\documentclass{sig-alternate-2013}

\newfont{\mycrnotice}{ptmr8t at 7pt}
\newfont{\myconfname}{ptmri8t at 7pt}
\let\crnotice\mycrnotice%
\let\confname\myconfname%

\permission{Permission to make digital or hard copies of all or part of this work for personal or classroom use is granted without fee provided that copies are not made or distributed for profit or commercial advantage and that copies bear this notice and the full citation on the first page. Copyrights for components of this work owned by others than ACM must be honored. Abstracting with credit is permitted. To copy otherwise, or republish, to post on servers or to redistribute to lists, requires prior specific permission and/or a fee. Request permissions from permissions@acm.org.}
\conferenceinfo{COSN'14,}{October 1 -- 2, 2014, Dublin, Ireland.}
\copyrightetc{Copyright 2014 ACM \the\acmcopyr}
\crdata{978-1-4503-3198-2/14/10\ ...\$15.00.\\
http://dx.doi.org/10.1145/2660460.2660469}

\usepackage{xcolor}
\usepackage{graphics}
\usepackage{times}
\usepackage{balance}
\usepackage{url}
\usepackage{subfigure}
\usepackage{graphicx}
\usepackage{epstopdf}
\usepackage{multicol}
\usepackage{amssymb}
\usepackage{pifont}
\newcommand{\cmark}{\ding{51}}%
\usepackage[sort,nocompress]{cite}

\usepackage{enumitem}

\usepackage{color}

\makeatletter
\def\@copyrightspace{\relax}
\makeatother

\begin{document}


\title{Partisan Sharing: \\ Facebook Evidence and Societal Consequences \\ \Large $[$Please cite the COSN'14 version of this paper$]$}

%
%
%
%
%

\numberofauthors{3} 
%
\author{
%
%
\alignauthor
Jisun An\\
       \affaddr{Qatar Computing Research Institute, Qatar}\\
       \email{jan@qf.org.qa}
\alignauthor
Daniele Quercia\\
       \affaddr{Yahoo Labs Barcelona, Spain}\\
       \email{dquercia@acm.org}
\alignauthor 
Jon Crowcroft\\
       \affaddr{University of Cambridge, UK}\\
       \email{Jon.Crowcroft@cl.cam.ac.uk}
}


\maketitle

\begin{abstract}
The hypothesis of selective exposure assumes that people seek out information that supports their views and eschew information that conflicts with their beliefs, and that has negative consequences on our society. Few researchers have recently found counter evidence of selective exposure in social media: users are exposed to politically diverse articles. No work has looked at what happens after exposure, particularly how individuals react to such exposure, though. Users might well be exposed to diverse articles but share only the partisan ones. To test this, we study \emph{partisan sharing} on Facebook: the tendency for users to predominantly share  like-minded news articles and avoid conflicting ones. We verified  four main hypotheses. That is, whether partisan sharing: 1) exists at all; 2) changes across individuals (e.g., depending on their interest in politics); 3) changes over time (e.g., around elections); and 4) changes depending on perceived importance of topics. We indeed find strong evidence for partisan sharing. To test whether it has any consequence in the real world, we built a web application for BBC viewers of a popular political program, resulting in a controlled experiment involving more than 70 individuals. Based on what they share and on survey data, we find that partisan sharing has negative consequences: distorted perception of reality. However, we do also find positive aspects of partisan sharing: it is associated with people who are more knowledgeable about politics and engage more with it as they are more likely to vote in the general elections.
\end{abstract}


\vspace*{-2mm}

\vspace*{-2mm}
\vspace*{-2mm}

\newpage 
\section{Introduction}
\noindent The media landscape affords people the opportunity to control which political messages they consume. With  freedom of choice comes responsibility, especially that of having a balanced news diet. Unfortunately, the opposite of a balanced diet -- selective exposure -- is likely to happen when consuming news. The theory of selective exposure holds that people tend to seek out political information confirming their beliefs and avoid challenging information. With a mix of experiments, surveys, and content analysis, decades of research have proved its existence across a variety of media  --  in newspapers, magazines, (cable) TV, radio, and online news sites~\cite{stroud2011,mutz2011,hahn2009,Lawrence2010}.

Selective exposure is thought to be highly problematic for democracy   --   for example, it is often associated with ``echo chambers''~\cite{sunstein2001}, whereby citizens befriend only like-minded others and do not talk to anyone else, resulting in segregated and polarized communities. Computer scientists have proposed different news aggregators that encourage politically diverse news consumption and try to mitigate the effect of selective exposure~\cite{blews@icwsm2008,jiang@sigir2008,resnick@chi2010,alice@icwsm2009,newscube@chi2009}.

The problem is that, based on previous work, we do not know what happens after exposure  --  how individuals react to it. People might be exposed to diverse articles but share only the partisan ones. Social media offers us a unique opportunity to study how people react upon exposure. We thus go beyond selective exposure and study \emph{partisan sharing} in an unobtrusive way and in large-scale: the tendency for users to predominantly \emph{share} (not only be exposed to) like-minded news articles and avoid conflicting ones.

Political scientists have been focusing on building a ``theory'' of news consumption and  they have done so  upon either \emph{self-reported} data of media \emph{consumption}  (and self-reporting can be inaccurate and error-laden~\cite{prior2009,Vavreck2007}), media \emph{selection} data (often generated from small-scale experiments) or \emph{actual} data of media \emph{exposure}, which does not necessarily translate into consumption  --  one might well be exposed to a TV show without paying too much attention.

Computer scientists, on the other hand, have been studying online news consumption for a while now~\cite{munson@icwsm2009,munson@icwsm2011,newscube@chi2009}. Individuals are increasingly turning to social networking sites to read and share political news, especially on Twitter and Facebook~\cite{an@icwsm2011, PewResearsh2012}. Researchers have been able to get hold of data on those sites and unobtrusively analyze sharing patterns of a large number of users during long periods of time. They have produced reliable data-driven analyses of sharing behavior without, however, focusing on the theoretical side. That is why hypothesis driven analyses, so common in political science, represent the next natural step for social media research.

\begin{figure}[t!] \begin{center}
\includegraphics[width=.47\textwidth]{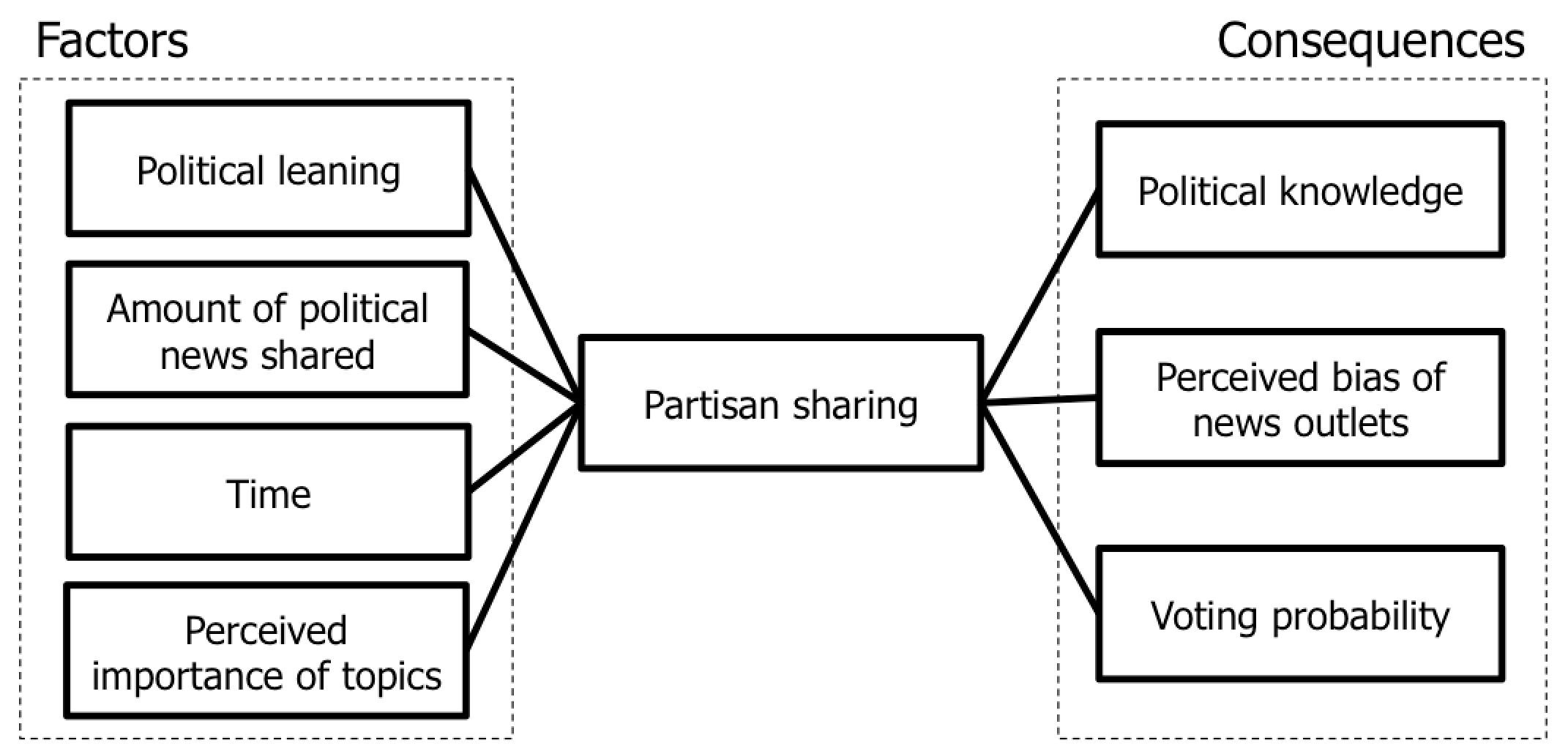}
\end{center} \vspace*{-6mm}
\caption{Factors and Consequences of Partisan Sharing.}
\vspace*{-1mm}\label{fig:framework} \end{figure}

To this end, we formulate a set of hypotheses from the political science literature (Figure~\ref{fig:framework}), analyze data from  Facebook (44,999 news articles from 37 popular US news sites and 12,495 Facebook user profiles) and gather evidence for or against partisan sharing. In so doing, we make the following main contributions:

\begin{itemize}
\item We derive a set of well-grounded and coherent hypotheses related to partisan sharing in social media from the literature in political science (Section~\ref{sect:se_framework}).  These hypotheses are about whether or not partisan sharing exists in social media; changes across individuals; changes over time; and impacts on society by being associated with specific people's political attitudes. 

\item We gather a representative US sample of Facebook users (Section~\ref{sect:methodology}) and test some of those hypotheses with that data (Section~\ref{sect:test}). We investigate the news media sources individuals share and we find that partisan sharing exists. Based on a measure of partisanship we will define, only 33\% of US Facebook news readers in our test sample can be considered moderate. Contrary to what the literature has posited, especially among conservatives, partisan sharing applies only to political news and disappears for non-political news (e.g., entertainment). We find that those who are partisan tend to be interested in politics. We have also been able to study how partisanship changes over time: we found that partisanship  among Facebook users within a US state tends to be stable, and changes only during primary elections to then go back to the original value.

\item Since Facebook data allows us to test most of the hypotheses but not all, we build a political site that recruits UK BBC viewers on Twitter and allows them to express their opinions about a weekly BBC political debate called Question Time (Section~\ref{sect:test_twitter}). Since Twitter differs from Facebook, we consider the hypotheses we have already tested on Facebook and find that they equally hold on the Twitter data. After this validity check, we test three new hypotheses regarding societal aspects of partisan sharing. We firstly find that people perceive a news outlet to be politically biased depending on their own political leanings (regardless of the objective bias) -- the farther their leaning from the outlet's, the more biased they perceive the outlet to be. This results in partisan individuals having a distorted perception of how biased news outlets are. However, partisanship has a positive aspect too. We indeed find that the more partisan individuals, the more politically knowledgeable they are (second hypothesis), and the more likely to participate to political life (e.g., to vote).

\end{itemize}

We brought data to bear on the phenomenon of partisan sharing. Contrary to popular belief, social media have  done little to broaden political discourse. Despite the political diversity social media have brought into one's news diet~\cite{an@icwsm2011}, individual news sharing has not been changed much contrary to traditional media. On one hand, social media users share news that matches their political beliefs and, as a result, may become increasingly divided; on the other hand, partisan sharing may encourage participation and news posting on social media sites. To mitigate the effect of partisan sharing, one might think of new ways of making sharing a bit less partisan and a bit more serendipitous, and this work offers an experimental basis for such future work.

\section{Related work}

\mbox{ } \\
\noindent \textbf{Selective exposure.} By analyzing news consumption on a variety of media (which included TV, radio, magazines, newspapers, online), Stroud~\cite{stroud2011} concluded that people tend to preferentially choose, read, and enjoy partisan news. A large body of literature shows supportive evidence for her findings~\cite{dilliplane2011,mutz2011,hahn2009,Lawrence2010,nie2010}. More recently, some researchers have reported situations in which selective exposure is lower than expected or totally missing. LaCour did not find any evidence for it in the  TV and radio consumption of 920 individuals in Chicago and New York~\cite{lacour2012}; Shapiro found an extremely low level of  it online~\cite{shapiro2011}; and An et al. even found that Twitter friends expand one's diversity of political news~\cite{an@icwsm2011}. 

However, despite its breadth, such a work and, for that matter, similar others suffer from the data under study: self-reported (and, as such, error-prone) data of news consumption. Starting from this criticism, LaCour directly measured how 920 individuals  from New York and Chicago have been exposed to news for 85 days~\cite{lacour2012}. These measurements were taken by cell phones that recorded participants' audio. He showed that self-reported data grossly overestimates exposure. It turns out that most people do not care much about politics and are thus on a meager news diet -- consequently, it does not really matter whether that diet is balanced or not. The problem is that audio-recording cell phones report what people are exposed to but not necessarily what they are paying attention to. A similar problem applies to An et al.'s work~\cite{an@icwsm2011}. The authors analyzed Twitter streams and found that Twitter friends  greatly expand one's  diversity of political news. However, it is not possible to quantify the  extent to which a Twitter user is actually paying attention to his/her own stream.

\mbox{ } \\
\textbf{Why it matters.} Selective exposure is often considered a threat to political and societal life. That is because it might influence people's beliefs in whom they should vote for~\cite{Mutz1996, Vigna2006}.  It also encourages  intolerance of dissent and results into ideological segregation~\cite{Bishop2008,PublicOpinion1999}. Adamic et al. ~\cite{adamic-blog@2005} showed that the political discourse on the blogsphere is not only partisan but also highly-polarized. The same pattern (i.e., polarized communities) has been observed in social media when looking at: usage of political hashtag ~\cite{conover@icwsm2011} and a distribution of tweets different stories receive~\cite{diego@cikm2013}. Political selective exposure also influences opinion formation on matters that have little to do with politics~\cite{stroud2011}. However, selective exposure is not always bad. It is also associated with political participation and political knowledge (i.e., factual information about politics)~\cite{stroud2011}: the more one is exposed, the more one is interested in politics and is knowledgeable about it.

\mbox{ } \\
\textbf{What to do.} Computer scientists have proposed different news aggregating systems that encourage politically diverse news consumption. For example, BLEWS~\cite{blews@icwsm2008} and NewsCube~\cite{newscube@chi2009} gather and visualize news articles on the same subject matter but with different political leanings, making people aware of the existence of media bias. Munson and Resnick have studied how different presentation techniques make politically diverse news articles more appealing than others. It turns out that making hostile news more appealing is quite challenging~\cite{resnick@chi2010}.

\mbox{ } \\
Considering the literature, we conclude that no work has studied one of the results of (selective) exposure: what people do after being exposed. Unlike traditional experiments, social media sites offers us a unique opportunity to observe sharing for a long period of time in unobtrusive ways. We set out to explore partisan sharing in the context of Facebook and Twitter, and we firstly do so by connecting sharing behavior to factors of selective exposure.

\section{Partisan Sharing}
\label{sect:se_framework}
\noindent To begin with, we define partisan sharing as the tendency for users to predominantly share like-minded news articles and avoid conflicting ones.\footnote{In this work, we only consider what people initially post (referred as `share') on social media.} Then, we derive five main hypotheses which were related to selective exposure in the literature and we now related to partisan sharing. That is, whether partisan sharing: 1) exists at all; 2) changes across individuals (i.e., depending on their interest in politics); 3) changes over time (i.e., around elections); 4) depends on perceived importance of topic; and 5) has any consequence in the real world (i.e., whether people vote or not).

\mbox{ }\\ 
\noindent \textbf{Existence of partisan sharing.} Its existence has been hitherto debated, as discussed in the previous section. Given that we will study news sharing in the context of Facebook and Twitter, our first hypothesis is: \emph{[H1] Individual's news sharing in social media is not balanced but suffers from partisan sharing.}\\

\noindent \textbf{Partisan sharing changes across individuals.} It has been found that those who have a settled tendency of reading about politics tend to seek out news confirming their political proclivities, mainly because they do not like to pay attention  to information challenging their views~\cite{stroud2011}. It is thus the case that ideological selectivity is predominant among partisan people and news junkies~\cite{Hahn2001,stroud2011}. Thus, our next hypothesis is: \emph{[H2] An individual's level of partisan sharing depends on}: \emph{[H2.1] political leaning}; and \emph{[H2.2] amount of political news shared}. Those three factors are listed, for convenience, in the block on the left in Figure~\ref{fig:framework}. That block collates factors associated with partisan sharing, which also include time, discussed next. \\

\begin{table}[t!] \begin{center} \small\frenchspacing
\begin{tabular}{l|cc} 
\hline
\textbf{Hypothesis} & \textbf{Facebook} & \textbf{Twitter} \\
\hline
\textbf{Dynamics} & &  \\
 \emph{[H1] Existence} &  \cmark & \cmark \\ 
 \emph{[H2.1] Political leaning} & \cmark & \cmark \\
 \emph{[H2.2] Amount of news shared} &  \cmark & \cmark \\
\emph{[H3] Changes over time} & \cmark   \\
\emph{[H4] Perceived importance}  & \cmark   \\
\hline
 \textbf{Societal consequences} & & \\
 \emph{[H5.1] Perceived bias}  &   & \cmark  \\
\emph{[H5.2] Political knowledge} &   & \cmark \\
\emph{[H5.3] Voting probability} &  & \cmark  \\
\end{tabular} \end{center} \vspace*{-4mm} \caption{List of Hypotheses.  \cmark indicates the hypothesis tested with Facebook/Twitter data. 
}
\label{tab:summary_hypothesis} \end{table}

\noindent \textbf{Partisan sharing changes over time.} People tend to pay more attention to politics during specific periods of time, for example, during elections when the attention to the political agenda is high. More generally we might hypothesize that \emph{[H3] Partisan sharing is highly prevalent in politically salient periods. }\\

\noindent \textbf{Partisan sharing changes depending on the perceived importance of certain topic.} People have their own preferences of  which  issues are important and ought to be in the political agenda. These preferences are often formed based on consumption of information from politically-biased outlets. Different outlets are associated with different topical priorities~\cite{shaw1972, pewsurvey2012}. Consequently, we hypothesize that \emph{[H4]. Partisan sharing is associated with one's perceived importance of certain topic.} \\

\noindent \textbf{Consequence of partisan sharing.} After having formulated hypotheses regarding the dynamics associated with partisan sharing, we now formulate three hypotheses regarding its societal consequences. Recent research has found that selective exposure results into polarization and societal fragmentation~\cite{stroud2011,sunstein2001}. Our next hypothesis is then \emph{[H5]. Partisan sharing is related to polarized political attitudes and, as such, affects one's:}
\begin{itemize}

\item[] \emph{[H5.1] Perceived political bias of news outlets.} Partisanship often influences perceptions, including perception of how biased a news outlet is. It has been shown that the same outlet is considered very differently by people depending on their political leanings~\cite{stroud2011}. A left-leaning outlet is perceived moderately (if not at all) biased by left-leaning people, while it is perceived highly biased by right-leaning people. 

\item[] \emph{[H5.2] Political knowledge.} Partisanship is associated with political knowledge: the more partisan, the more knowledgeable about politics ~\cite{stroud2011}.

\item[] \emph{[H5.3] Voting probability.} When people decide for whom to vote, they again rely on partisan media, which exert considerable influence~\cite{Mutz1996, Vigna2006}. 
\end{itemize}

Having a list of factors and of consequences  associated with partisan sharing, we are now ready to test their importance and we do so by using two datasets: Facebook and Twitter datasets. Table~\ref{tab:summary_hypothesis} summarizes the hypotheses and report which dataset is used to test which hypothesis. We mainly use Facebook data for the analysis, and we validate and complement  results with Twitter data (Section~\ref{sect:test_twitter}).

\section{Methodology}
\label{sect:methodology}

\noindent To begin with, we perform the following steps: gather a Facebook dataset of news sharing, consider only articles about politics, determine the media slant of the corresponding outlets, and compute each user's partisan skew. The higher an individual's partisan skew score, the higher his/her partisan sharing.

\mbox{} \\
\textbf{Facebook news sharing.}  More than five million Facebook users have been able to take a variety of genuine personality and ability tests by installing an application called myPersonality.\footnote{http://www.mypersonality.org/wiki} Users can also opt in and give their consent to share their personality scores and profile information, and 40\% have chosen to do so. We gather news sharing information from a random subset of those users: 228,064 Facebook users who shared (i.e., posted) 4.9M links. 

To avoid temporal biases generated by the adoption of the Facebook application, we focus on a specific period of time that is stable enough: from April to September 2010. During that period, we gathered 44,999 articles posted by 12,495 users: 37\% of links are successfully classified as political news articles through Alchemy API (discussed shortly). The articles they post come from 37 news sites (a few representative news sites are displayed in Table~\ref{tab:url-domains}.\footnote{6 news outlets are considered to be at the center of the political spectrum: CNN (13,753 (39.7\%)), NewsWeek (1,101 (45.8\%)), Arizona Central (363 (37.2\%)), The Atlantic (384 (44.2\%)), PBS (808 (51.2\%)), and Christian Science Monitor (603 (31.2\%))}). The demographic of these users reflects the general Facebook population in USA \footnote{Ugander et al. have reported that the age of 140M USA Facebook users ranges from 13 to 60+ where 20s and 30s are the dominant. Also the median number of social contacts is 100, yet, the distribution is highly skewed (there are few people having more than 1000 social contacts), resulting in 190 as an average value~\cite{ugander@corr2011}.}: their number of social contacts is between 30 and 1000 and whose age is comprised between 18 and 54. This group is composed of 7,372 women (59\%) and 5,037 men (41\%) with a median age of 23. Table~\ref{tab:summary_user_detail} reports the demographic details of Facebook users in our dataset. 

Each user has posted 2.85 articles on average; while 66.5\% of users have posted only one article. Based on their activity level (i.e., number of news shared), we find 4,710 users who posted more than two news articles, 2,113 users who posted more than four news articles, and 950 users who posted more than eight news articles on Facebook. We confirm that demographic information of those groups are consistent with that of general Facebook population.

\begin{table}
\begin{center}
\small \frenchspacing
\begin{tabular}{lr|lr}
\hline
\multicolumn{4}{c}{\textbf{Facebook}} \\
\hline
\multicolumn{2}{c}{\textbf{Liberal}} & \multicolumn{2}{c}{\textbf{Conservative}}\\
\hline
    huffingtonpost &11,236 (45.2\%) & foxnews & 3,774 (51.5\%)\\
    nytimes &10,083 (41.3\%) & online.wsj &2,767 (54.1\%) \\
    msnbc.msn& 4,892 (30.7\%) & nydailynews &1,433 (26.8\%) \\
    abcnews.go & 2,752 (23.2\%) & nypost& 695 (31.9\%) \\
    washingtonpost & 2,468 (58.3\%) &  politico & 670 (87.3\%)   \\
    time & 2,104 (34.9\%)  &  forbes & 565 (20.7\%)\\
    cbsnews & 1,868 (22.8\%)  &   washingtontimes &432 (73.4\%) \\
   bostonglobe& 1,260 (29.3\%)  & townhall & 426 (91.3\%)   \\ 
    latimes & 1,239 (37.3\%)  & nationalreview & 288 (65.9\%) \\
     salon & 967 (64.8\%)  &  chicagotribune & 239 (36.4\%)  \\
    slate & 923 (44.9\%) & bostonherald & 118 (50.1\%)  \\
     sfgate & 779 (34.1\%) &  usnews& 130 (36.2\%)  \\
    wnd &441 (72.1\%)  & newsmax &93 (76.6\%)  \\
  newyorker & 368 (33.9\%)  &  weeklystandard &60 (79.5\%)  \\
\end{tabular}
\end{center}
\caption{List of the News Sites Shared by Facebook Users. We remove ``.com'' from domain name of news sites. Each outlet comes with its total number and proportion of political news articles.}
\label{tab:url-domains}
\end{table}

\mbox{} \\
\textbf{Only news articles about politics.}\label{sect:methodology} To select only the articles about politics, we need to be able to classify articles into categories, and select those that fall into politics. To that end, we use Alchemy text classification API.\footnote{http://www.alchemyapi.com} We use this API because it has been shown that it entails superior classification performance compared to other popular classifiers~\cite{quercia@websci2012}. Alchemy API is a suite of natural language processing tools. It is capable of assigning a plain English category to any given string of text (a tweet, for instance), along with a certainty score from 0.0 to 1.0, which represents the API's degree of belief that the text pertains to that category. It can also take a URL as an argument -- it then classifies the textual content of the document it finds there. Alchemy can choose from the following 12 topics: Arts Entertainment, Business, Computer Internet, Culture Politics, Gaming, Health, Law Crime, Recreation, Religion, Science Technology, Sports, and Weather. Hence we are using Alchemy topics as a gold standard in our work. We excluded 5,705 URLs that are categorized as ``None'' (e.g., broken link) and URLs that have low confidence values ($<$ 0.5 on Alchemy's scale of $[0,1]$). Then we take the remaining ones that are classified under ``Culture/Politics'' for our analysis (16,729 news articles).

\mbox{} \\
\textbf{Determining media slant.}  We need to classify news outlets into liberal, conservative, or center. Since the outlets in the Facebook dataset are mainly in US (Table~\ref{tab:url-domains}), we consider four classification schemes previously used in the literature of political science: 1) \textit{Left-Right}, which classifies a large number of news outlets, including those with online presence only\footnote{http://left-right.us}; 2) \textit{MondoTimes}, which classifies news outlets based on user votes on a crowdsourcing platform\footnote{http://mondotimes.com} (this has been used in Gentzkow and Shapiro's work~\cite{shapiro2010}); 3) \textit{Gentzkow and Shapiro}'s novel classification, which relies on term similarity between political speeches and news articles~\cite{shapiro2010}; and 4) \textit{Larcinese} et al.'s classification, which relies on the amount of coverage media outlets give to U.S. political scandals~\cite{Larcinese2010}. The results reported later on do not change depending on which of the four classification we use. That is largely because the four schemes show  high agreement, as we shall discuss in Section~\ref{sect:discussion}. Thus we report the results only for the first classification.

\begin{table}[t!] \begin{center} \small\frenchspacing
\begin{tabular}{r|l} \hline
\multicolumn{2}{c}{\textbf{Facebook}} \\
\hline
  \textbf{Age} &$<$20(25\%), $<$30(31\%), $<$40(18\%), $\geq$40(16\%) \\
  \textbf{Gender} & Male (41\%), Female (59\%) \\
  \textbf{Partisanship} & Partisan (59.6\% (Lib (69.9\%),Con (30.1\%))), \\
  & Non-partisan (40.4\%) \\
   
\end{tabular} \end{center}  \caption{Details for our 12,855 Facebook Users.}
\vspace*{-4mm}
\label{tab:summary_user_detail} \end{table}

\begin{figure*}[ht!] \begin{center}
\subfigure[Political news (All)]{\includegraphics[width=.23\textwidth]{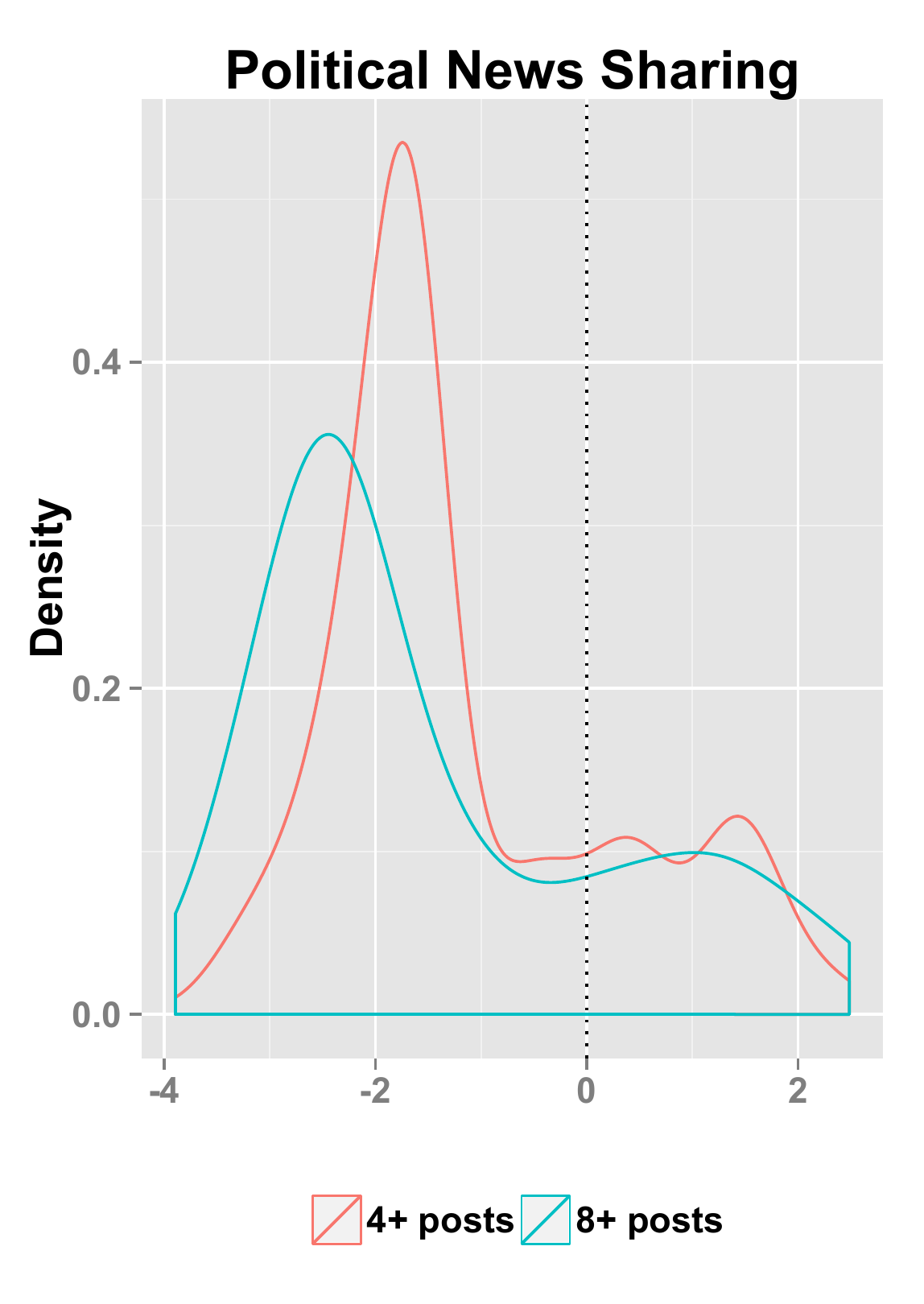} \label{fig:density_politics_narticleAll} }
\subfigure[Political news (Partisan)]{\includegraphics[width=.23\textwidth]{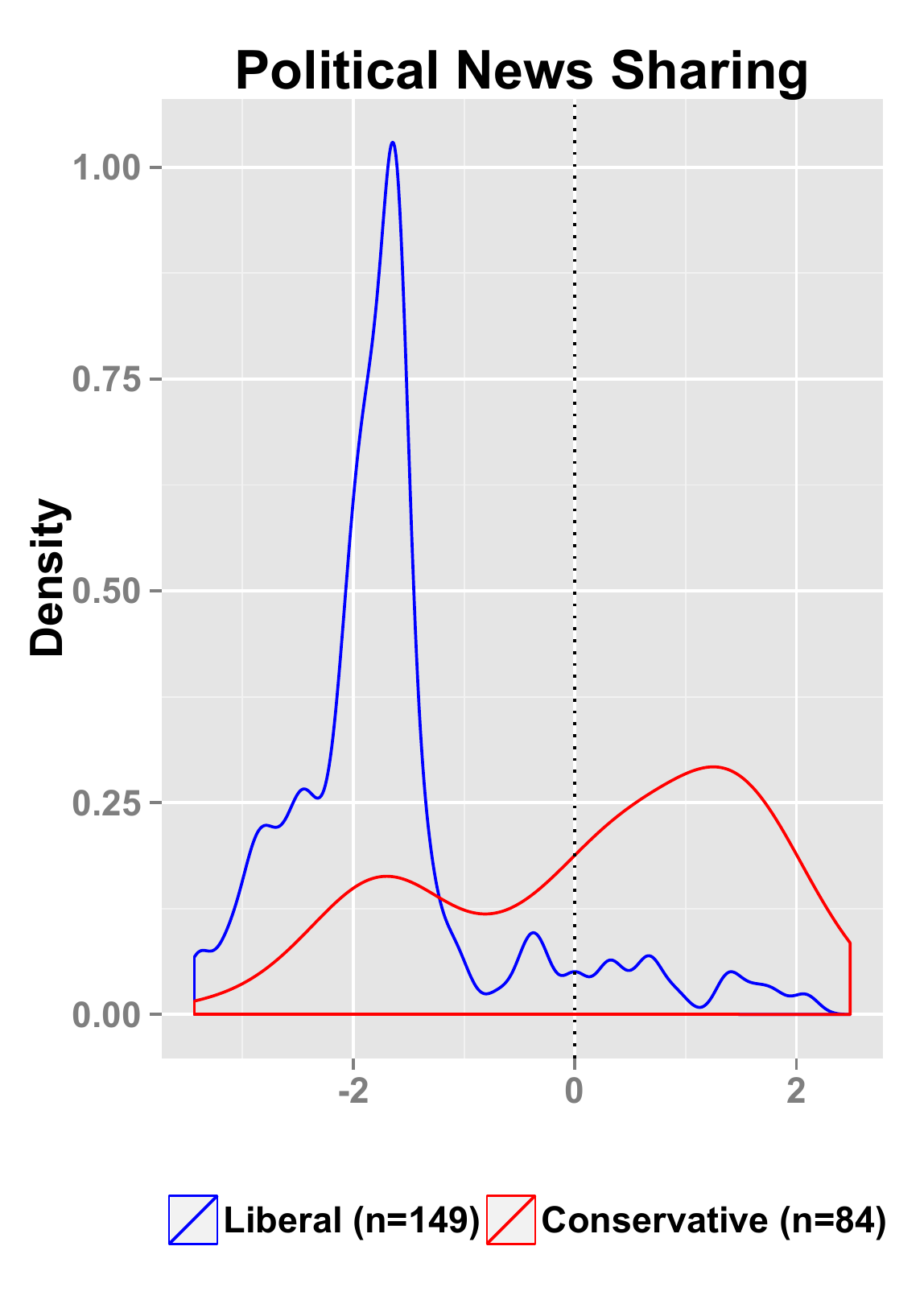} \label{fig:density_politics_narticleAll_partisan}}
\subfigure[Non-political news (All)]{\includegraphics[width=.23\textwidth]{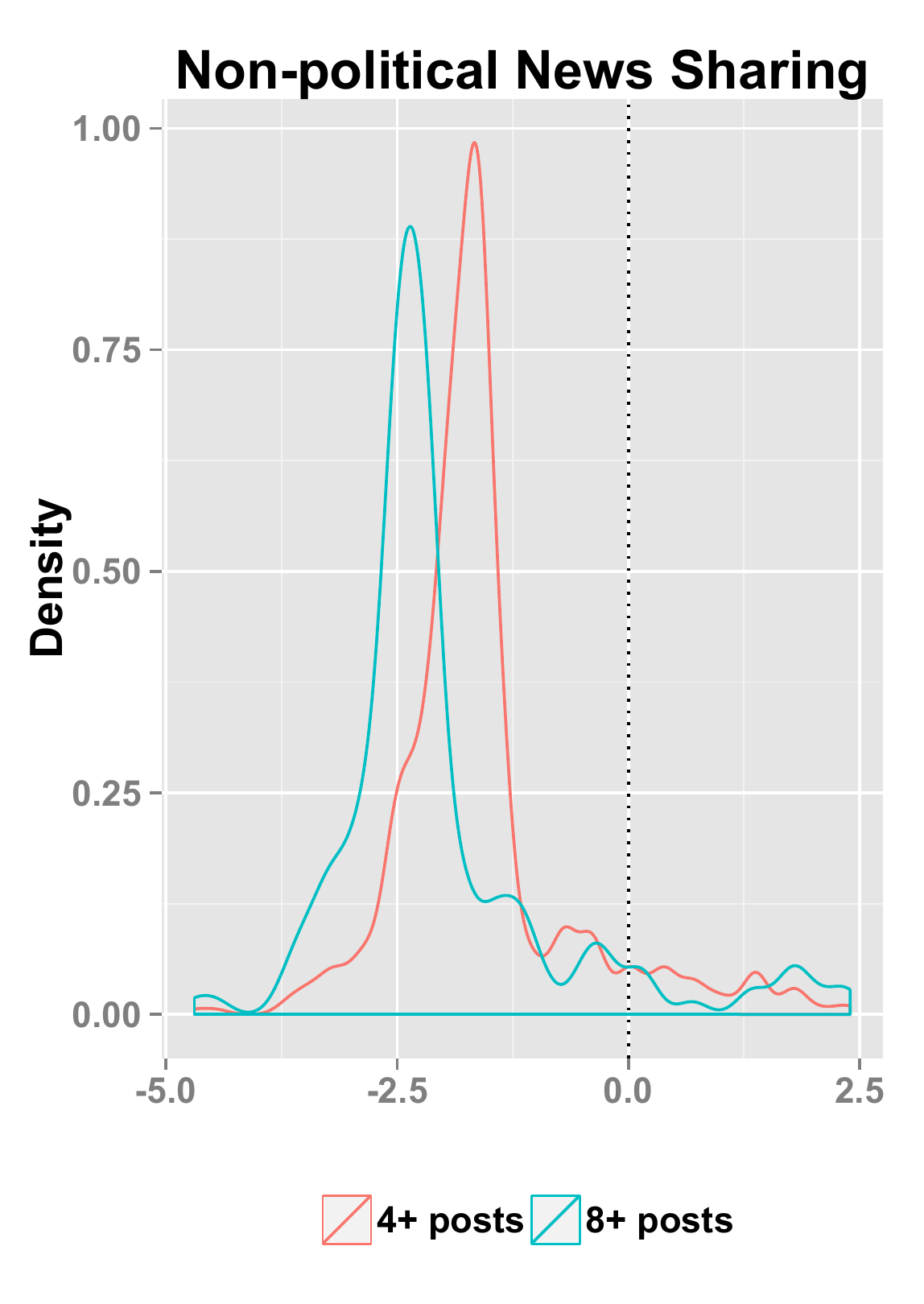} \label{fig:density_other_narticleAll} }
\subfigure[Non-political news (Partisan)]{\includegraphics[width=.23\textwidth]{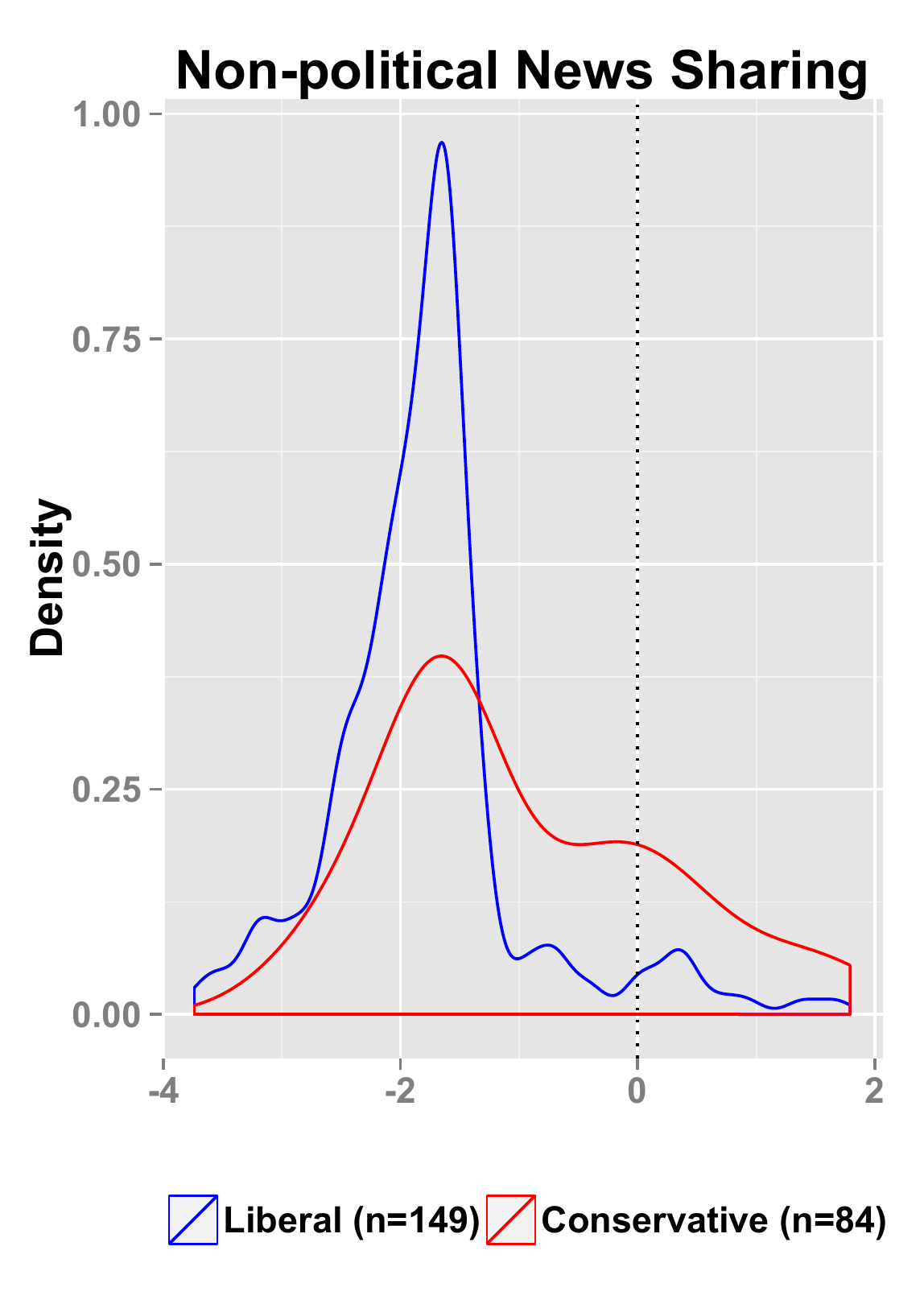} \label{fig:density_other_narticleAll_partisan}}
\end{center} \vspace*{-6mm}
\caption{\textbf{Net Partisan Skew of political news by} \emph{(a)} \emph{Activity Level}. For any of the two activity levels (4+ articles and 8+ articles posted), there are two peaks reflecting liberal views (left peak) and conservative views (smaller right peak), respectively. The majority of our Facebook users are liberal.  \emph{(b)} \emph{Party}. We consider users who share 4+ political articles. Liberal users share more liberal news outlets and are centered around a net partisan skew of -1.8, while conservative users share more conservative news and are centered around 1.4. \textbf{Net Partisan Skew of soft news (e.g., Arts Entertainment) by} \emph{(c)} \emph{Activity Level}. For any of the two activity levels (4+ articles and 8+ articles posted), the distribution is unimodal and centered around -1.7 and -2.43, respectively. Both Liberal and Conservative are sharing soft news coming from liberal outlets. \emph{(d)} \emph{Party}. We consider users who share 4+ political articles. Liberal users post more liberal news outlets and are thus centered around a net partisan skew of -1.8.} 
\vspace*{-1mm}\label{fig:kdf_net_skew_politics} \end{figure*}

\mbox{ } \\
\textbf{Measuring partisanship: Net partisan skew.} To measure partisan sharing, in line with previous work~\cite{lacour2012}, we focus on news posting from partisan sources (i.e., those that are classified as either conservative or liberal) and compute the \emph{net partisan skew} as the number of conservative news postings minus that of liberal news (news counts, being  skewed, undergo a logarithm transformation):

\begin{equation}
leaningScore_{conservative} - leaningScore_{liberal}
\label{formula:net_partisan_skew}
\end{equation}

\begin{equation}
leaningScore_{p} = ln(\# \textrm{news articles of $p$})
\end{equation}

\noindent $leaningScore_{p}$ is 0 when the news count of $p$ is 0. The partisan skew is our main measure of partisan sharing and reflects how balanced a user's news sharing is -- for example, it is zero if the user posts an equal amount of conservative and liberal news; it is  $ \pm$ 1 if, for every 2.7 ($\approx e^{1}$) conservative (liberal) articles, the user posts 1 liberal (conservative) article; and it is $ \pm$ 2 if, for every 7.4 ($\approx e^{2}$) conservative (liberal) articles, the user posts 1 liberal (conservative) article. We use this metric to be able to compare our result to that of previous work~\cite{lacour2012}, which has incidentally shown limited evidence of selective exposure.

\section{Partisan Sharing in Facebook}
\label{sect:test} 

\noindent We will test to which extent partisan sharing exists (Section~\ref{sect:existence}) and how it changes  across individuals (Section~\ref{sect:se_individual}) and over time (Section~\ref{sect:se_time}), we will then study its association with perceived importance of topics (Section~\ref{sect:importance}). Table~\ref{tab:summary_hypothesis} reports which hypotheses has been tested upon Facebook dataset.


\newpage
\subsection{Existence}
\label{sect:existence}

\noindent \emph{\textbf{[H1]} Individuals' news sharing in social media is not balanced but suffers from partisan sharing. }

To measure the extent to which partisan sharing exists, we analyze news sharing for articles coming from partisan news outlets -- that is, from outlets that can be labeled as either conservative or liberal. 
We compute net partisan skew using expression~(\ref{formula:net_partisan_skew}) (how balanced one's news sharing is): a positive score represents users sharing news from conservative outlets, while a negative one indicates sharing from liberal outlets. 
The theory of partisan sharing suggests that we should find a binomial distribution, with conservative users sharing predominantly conservative news articles, and liberal users sharing predominantly liberal ones. 
Figure~\ref{fig:density_politics_narticleAll} displays the distribution of net partisan skew in the form of kernel density estimates for two sets of users -- low-activity users who posted at least 4 articles
and high-activity ones who posted more than 8 articles. Each curve shows two peaks, reflecting two user segments -- one sharing exclusively liberal news, and the other sharing exclusively conservative news. This is true for both curves, suggesting that partisan sharing holds not only for high-activity users  but also for low-activity ones. 
Based on self-reported political affiliations, clearly denoted as either ``liberal'' or ``conservative'', on Facebook, we separate liberal users ($N$ = 149) from conservative ones ($N$=84) and compute their partisan skew (Figure~\ref{fig:density_politics_narticleAll_partisan}). 
We find that all of them share a considerable number of like-minded news and systematically avoid counter-attitudinal news. 
The majority of liberals (conservatives) read one counter-attitudinal article every 6 (4) like-minded articles.

Finally, since it has been shown that selective exposure to political news ends up influencing news sharing on matters that are not strictly related to politics~\cite{stroud2011}, we analyze sharing of not only political news but of any type of news. 
We find that, when sharing news about, say, Arts Entertainment, people do not constraint their general (non-political) news sharing to outlets matching their political beliefs, resulting in a unimodal (skewed to the left) distribution (Figure~\ref{fig:density_other_narticleAll}). 
Surprisingly, conservatives tend to share non-political news from liberal outlets (Figure~\ref{fig:density_other_narticleAll_partisan}).

\subsection{Changes across individuals}
\label{sect:se_individual}
\noindent Despite the evidence that partisan sharing occurs, it is clear that not everyone shares like-minded news to the same extent -- after all, 32.8\% of users have a net partisan skew in the range as low as [-1,1]. Does partisan sharing change depending on users' characteristics? Previous studies in political science have posited that \emph{an individual's level of partisan sharing depends on:} \emph{\textbf{[H2.1]} political leaning}; and \emph{\textbf{[H2.2]} amount of news sharing}.

\begin{figure}[th!] \begin{center}
\subfigure[US Facebook]{ \includegraphics[width=.185\textwidth]{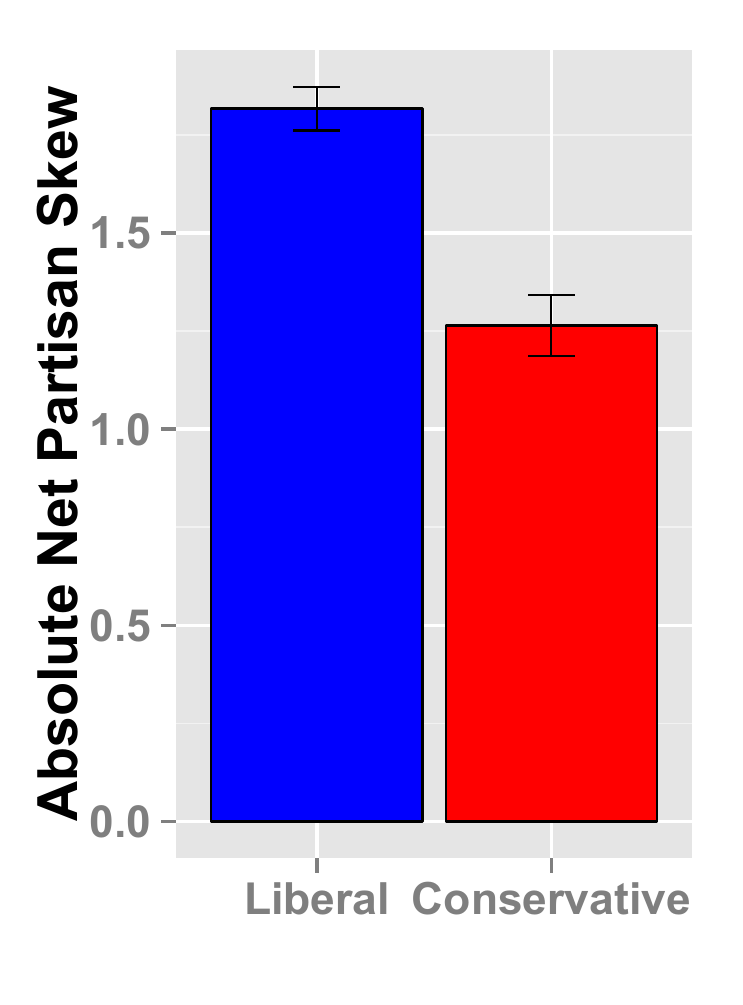} \label{fig:bar_netskew_partisan}}
\subfigure[US Facebook]{\includegraphics[width=.255\textwidth]{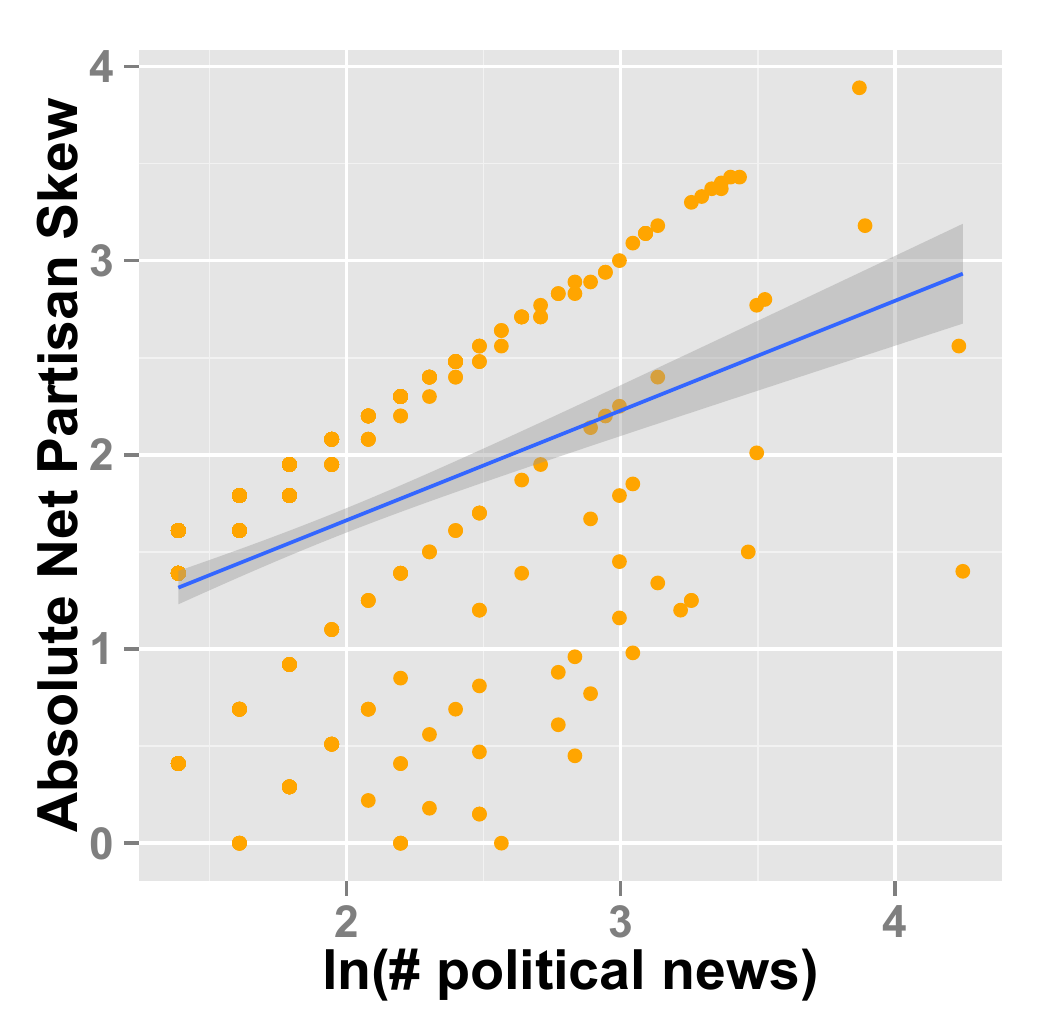} \label{fig:dot_netskew_numpoliticalnews} }
\end{center} \vspace*{-6mm}
\caption{Net Partisan Skew by \emph{(a)} Party and \emph{(b)} Activity. The absolute value of partisan skew is for US Facebook users who have shared 4+ articles.}
\vspace*{-4mm}\label{fig:netskew} \end{figure}

We test whether partisan sharing changes depending on one's political leaning (\emph{H2.1}). For a fair comparison, we first check whether the activity level of liberals and conservative are comparable. We find that they are similar -- on average liberals share 9 news articles, while it is 8.2 news articles shared for conservative. Also unpaired t-test confirms that there is no difference on their activity level (the hypotheses was rejected).
Then, we plot the (absolute value) of partisan skew for conservative and liberals (Figure~\ref{fig:bar_netskew_partisan}). 
By running unpaired t-test on these values, we find that liberals tend to be more partisan (with net skew of 1.82) than conservative (with net skew of 1.26) ($t$(166.535) = 5.805,  $p<0.0005$).
For every counter-attitudinal news article shared, liberals will also share 6.2 like-minded articles, while conservative will only share 3.5 like-minded articles. 
This means that conservative users are less polarized than liberal ones, sharing 43\% less like-minded articles. This is in line with work by LaCour who found that ``Democrats, as a group, watch slightly more like-minded news, while on average Republicans have a more balanced media diet''~\cite{lacour2012}.

Next, to test \emph{H2.2}, we plot the net partisan skew (absolute value) against news sharing (i.e., the logarithm of number of shared news articles) in Figure~\ref{fig:dot_netskew_numpoliticalnews}. The higher the news sharing, the higher the  partisan skew. For example, users who shared 4 articles ($\approx$  1.2 on the \emph{x}-axis) have an average partisan skew of 1.2: for every 3.3 ($\approx e^{1.2}$) conservative (liberal) articles, those users share 1 liberal (conservative) article. Higher-level activity users, say, those who shared 20 articles ($\approx$  3 on the \emph{x}-axis) have partisan skew of 2: for every 7.4 ($\approx e^{2}$) like-minded articles, those users share only 1 counter-attitudinal article. As users share more news, they also share more partisan news. For statistical test, we run two popular correlation with a Pearson's correlation coefficients of $r=.41$ ($p<$0.0005) and with a Spearman's correlation coefficients of $r=.46$ ($p<$0.0005). This finding runs contrary to what recent work has found~\cite{lacour2012} and confirms the partisan sharing hypothesis: as news sharing increases, readers tune out the other side.

\subsection{Changes over time}
\label{sect:se_time}

\noindent From February to September 2010, primary elections were held in the USA (elections in which each political party nominates candidates for an upcoming general election), and different States held them in different months (e.g., Indiana in May, California in June).\footnote{http://www.bbc.co.uk/news/world-us+canada-10634453} We consider the time window in which  our Facebook data overlaps with the election period and obtain news sharing data for two sets -- States that voted in May (10 states out of 10 that held elections) and those that voted in June (12 out of 13). We then test the following hypothesis:

\noindent \emph{\textbf{[H3]} Partisan sharing is  prevalent in  politically salient periods (e.g.,  during elections). }

\begin{figure}[t!] \begin{center}
\includegraphics[width=.49\textwidth]{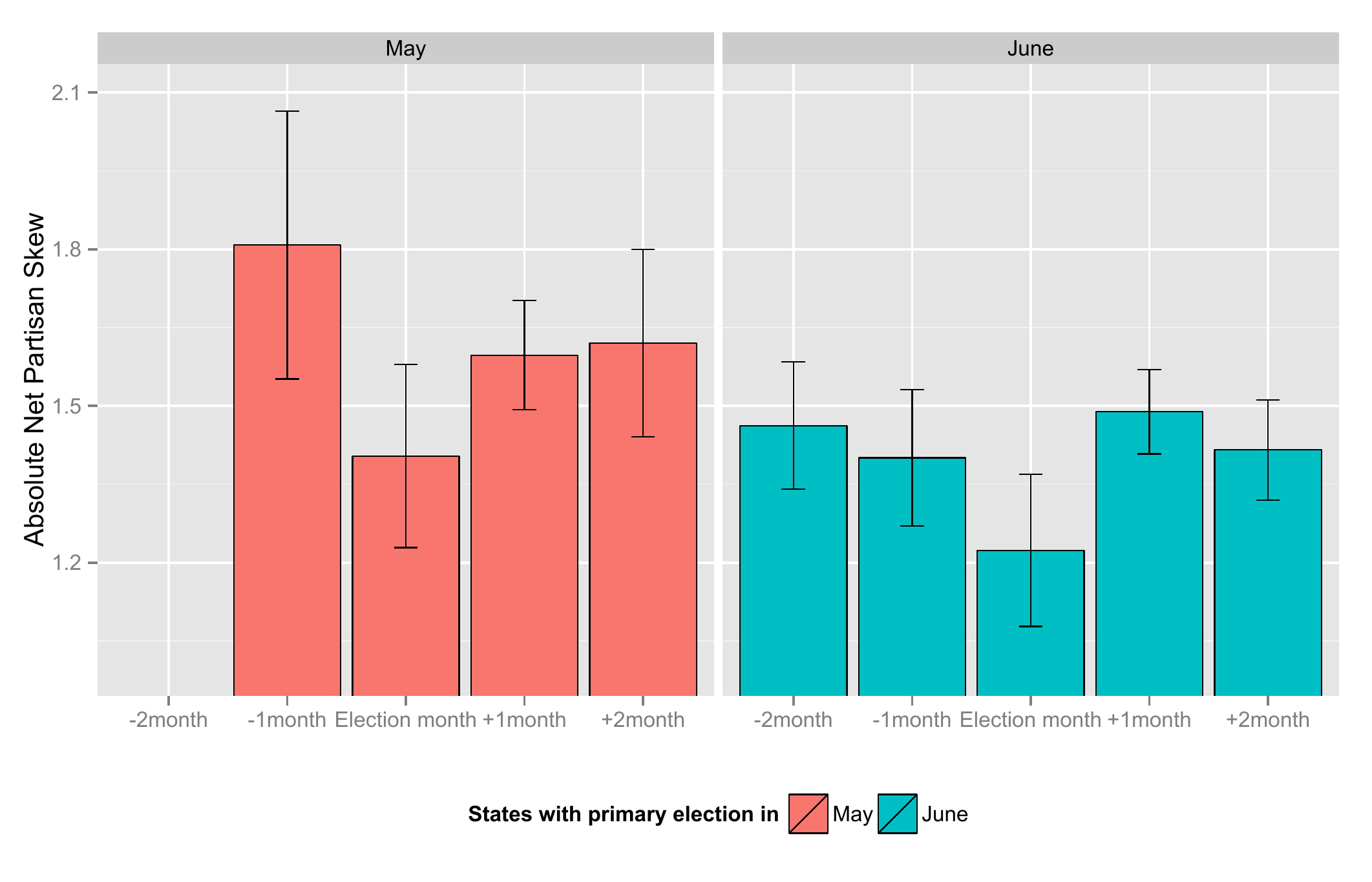}
\end{center} \vspace*{-6mm}
\caption{Absolute Net Partisan Skew by voting US states. That had primary elections in May (first barplot), and those in June (second barplot).}
\vspace*{-1mm}\label{fig:bar_primary_election} \end{figure}

Figure~\ref{fig:bar_primary_election} shows the average partisan skew (absolute value) for the two sets -- for both of them, partisanship is minimum  in the election month and tends to \emph{increase} to a stable point outside that period. For the States voting in May, the absolute average partisan skew is 1.4 in the election month and is around 1.6 outside it. This means that, during elections, for every 4 ($\approx e^{1.4}$) conservative (liberal) articles, users post 1 liberal (conservative) article. Outside elections, partisan skew increases:  users need to post 5 ($\approx e^{1.6}$) conservative (liberal) articles to then post 1 liberal (conservative) article. The same pattern holds for the States voting in June, where the ratios are 3.3-to-1 during elections and 4-to-1 outside them. Contrary to our expectation, this result (i.e., minimum partisanship during elections) seems to suggest that Facebook users tend to make their news diets both richer and more balanced during elections. Or it might well be that partisans share news articles from hostile outlets, just to make fun of them. If we were to have comments associated with the act of sharing, we would have studied their sentiment. Unfortunately, we do not have such data. However, we should stress that temporal evolution of partisanship has never been studied before, and this result, albeit preliminary, suggests that it is a research direction that ought to be in the agenda.

\subsection{Changes depending on perceived importance}
\label{sect:importance}

\noindent \textit{[H4]} Partisan sharing is associated with one's perceived importance of certain topic.

To test \emph{H4}, i.e., whether media news supply is tailored to partisan consumption (e.g., whether Fox News tailors its offering to its partisan readers as opposed to moderate ones),  we identify two different classes of users -- those who are partisan (high net skew) and those who are moderate (low net skew) -- and see whether they consume different topics to a different extent. To do so, we match (moderate and partisan) conservative  users with conservative outlets, and (moderate and partisan) liberal ones with liberal outlets. We then compute the total supply-demand divergence for four subgroups of the form user\&outlet: partisan\&conservatives, moderate\& conservatives,  partisan\&liberals, and moderate\&liberals.  The divergence for any of the four sets is then:

\begin{eqnarray}
\textrm{divergence}= \sum_\textrm{topic} |\textrm{demand}_\textrm{topic} -   \textrm{supply}_\textrm{topic} | 
\end{eqnarray}

computed over all topics covered by Alchemy API (Section~\ref{sect:methodology}), $\textrm{demand}_\textrm{topic}$ is the proportion of news articles a certain type of users (e.g., moderate) have consumed in  that $\textrm{topic}$, and $\textrm{supply}_\textrm{topic}$ is the proportion of news articles the partisan media have supplied for that $\textrm{topic}$. If an outlet supplies articles consumed by a certain class of users (e.g., partisan vs. moderate), then the divergence is zero for that class. By contrast, it is highest when the supplied articles do not meet the demand of that class of users at all. We expect that supply matches the partisans' news demand (lower divergence) rather than the moderates' (higher divergence). News outlets tend to meet the demand of partisan users twice as much as (1.7 $x$) the demand of moderate ones.

\subsection{Summary}

\noindent We have found that partisan sharing: 1) exists, but contrary to the literature, such selectivity is limited to political news; 2) changes across individuals -- people who are interested in politics tend to have stronger partisanship; 3) changes over time, in particular, their political diversity increases during the election period; and 4) is associated with perceived importance of topic -- news outlets match the information needs of partisan rather than moderate online readers.

\section{Partisan Sharing in Twitter}
\label{sect:test_twitter}

\noindent We first analyzed a Facebook dataset of shared news articles. Since we can test only part of the hypotheses on Facebook data, we also build a political engagement site connected to Twitter to perform the remaining hypotheses. As shown in Table~\ref{tab:summary_hypothesis}, we validate the first three hypotheses and we newly investigate the last three hypotheses with a Twitter dataset. 

\begin{figure}[t!] \begin{center}
\includegraphics[width=.45\textwidth]{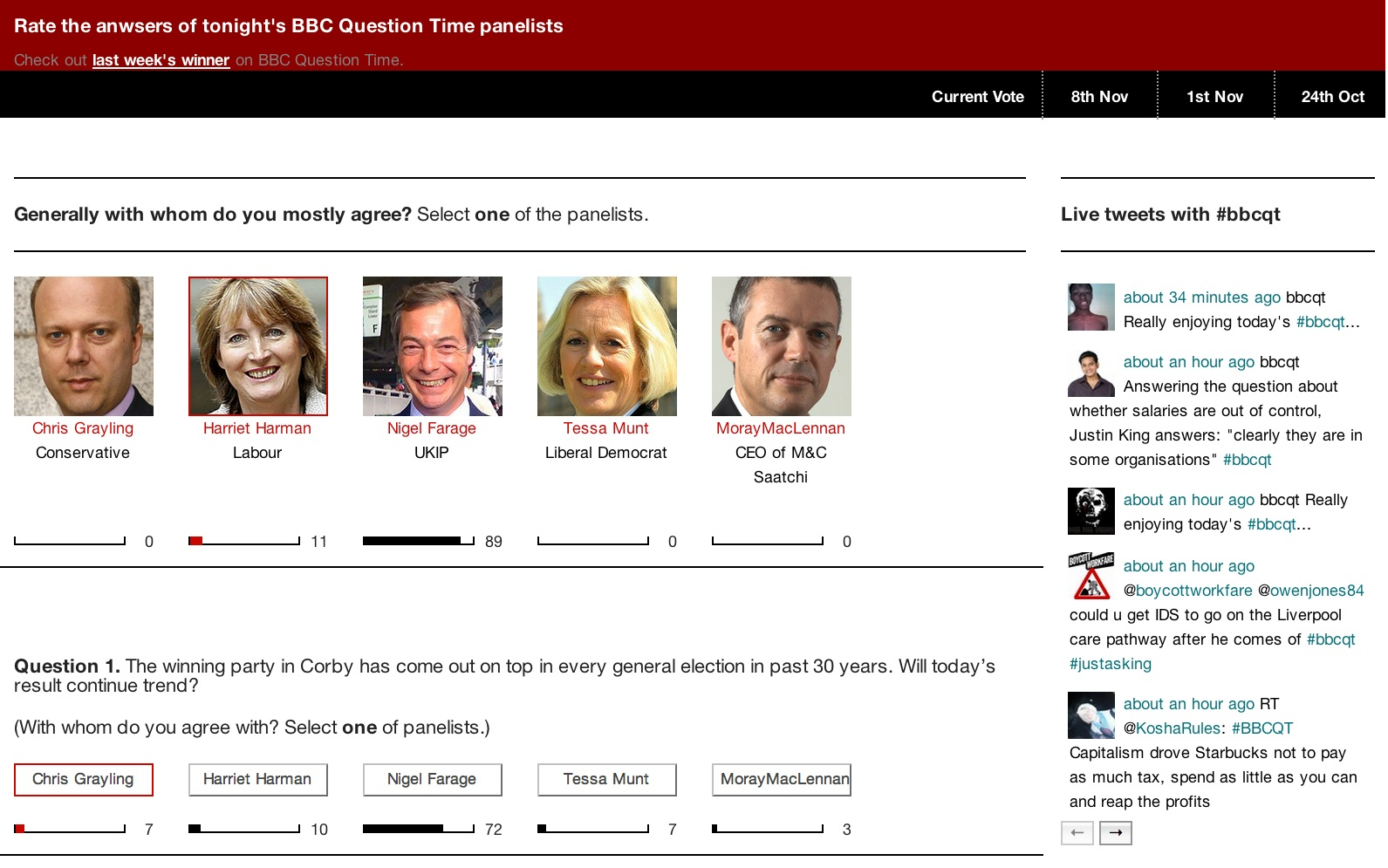}
\end{center} \vspace*{-6mm}
\caption{Screenshot of our application.}
\vspace*{-1mm}\label{fig:votingtime_screenshot} \end{figure}

\subsection{Methodology}

\noindent To conduct the study, we follow steps similar to those described in Section~\ref{sect:methodology}: gather a Twitter dataset, extract political news articles, determine media slant, and measure net partisan skew of each user.

\mbox{ } \\
\textbf{Twitter news sharing of BBC audience.} Question Time is a topical debate television program in UK. The show typically features five panelists -- three UK politicians from the major parties plus two public figures -- who answer pre-selected questions. On 24 September 2009, the show launched its Twitter presence  and, by June 2011, it became one of the most-tweeted shows in the UK, with more than 5K tweets using the \#bbcqt hashtag during each week. We have implemented a web application\footnote{http://www.votingtime.org.uk} with which BBC viewers  could select the panelist they found more convincing (see Figure~\ref{fig:votingtime_screenshot}). These viewers were recruited by posting messages on the Twitter stream of \#bbcqt hashtag. We made the application available during three weekly programs in 2012 (Oct 24, Nov 1, and Nov 7). During this period, we had 102 users who voted, among whom 71 reported their Twitter usernames and 35 answered survey questions (e.g., \textit{which party is closest to your political preference?, how politically biased are the following news outlets? did you vote in the last general election?}). Table~\ref{tab:summary_user_detail_uk} reports details of those user who responded. For the 71 users who reported their valid usernames, we crawled their tweets. We gathered 1,008 political news articles posted by the 71 users for the last 5 months of data collection, which happened to come from 10 different UK news sites reported in Table~\ref{tab:url-domains}.\footnote{BBC News (375 (48\%)) was one of the popular news sources among our users, and it is known to be neutral.}  The average number of articles per individual is 14 and the median is 10.

\mbox{} \\
\textbf{Determining media slant.}
In the Twitter dataset, we have mainly UK outlets, for which the literature does not offer any classification. We thus contacted  three UK political journalists and ask them to classify the outlets into liberal, conservative, or neutral for us (Table~\ref{tab:url-domains-uk}). We measure the inter-rater agreement using Cohen's kappa coefficient~\cite{cohen1960}, which results in 1 if two raters are in complete agreement. The overall kappa score among them on the 10 UK news sites was as high as 0.918. Under majority rule, we take a political leaning for a media source that is preferred by a majority of journalists.

\begin{table}[t!] \begin{center} \small\frenchspacing
\begin{tabular}{r|l} \hline
\multicolumn{2}{c}{\textbf{Twitter}} \\

\hline
 \textbf{Age} & $<$20 (8\%), $<$30 (7\%), $<$40 (33\%), $<$50 (22\%), $\geq$ 50 (30\%) \\
  \textbf{Gender} & Male (65\%), Female (35\%) \\
  \textbf{Partisanship} & Partisan (81.2\% (Labour (40\%), Lib(17\%), \\
   & Ind (31\%), Con (12\%))), Non-partisan (18.8\%) \\
\end{tabular} \end{center} \vspace*{-4mm} \caption{Details for our 35 Twitter Users interviewed.}
\vspace*{-2mm}
\label{tab:summary_user_detail_uk} \end{table}

\begin{table}[t!]
\begin{center}
\small \frenchspacing
\hspace*{-5mm}
\begin{tabular}{lr|lr}
  \hline
  \multicolumn{4}{c}{\textbf{Twitter}} \\
 \hline
  \multicolumn{2}{c}{\textbf{Liberal}} & \multicolumn{2}{c}{\textbf{Conservative}}\\
  \hline
  guardian & 482 (40.6\%)&  telegraph& 350 (45.7\%)\\
independent& 141 (37.4\%)&   dailymail & 149 (31.9\%)\\
 mirror & 67 (34.1\%)& thesun& 64(36.4\%)\\
\end{tabular}
\end{center}
\vspace*{-3mm}
\caption{List of the News Sites Shared by Twitter Users. We remove ``.co.uk'' from domain name of news sites.
}
\vspace*{-4mm}
\label{tab:url-domains-uk}
\end{table}

\begin{figure*}[th!] \begin{center}
\subfigure[Political news (All)]{\includegraphics[width=.23\textwidth]{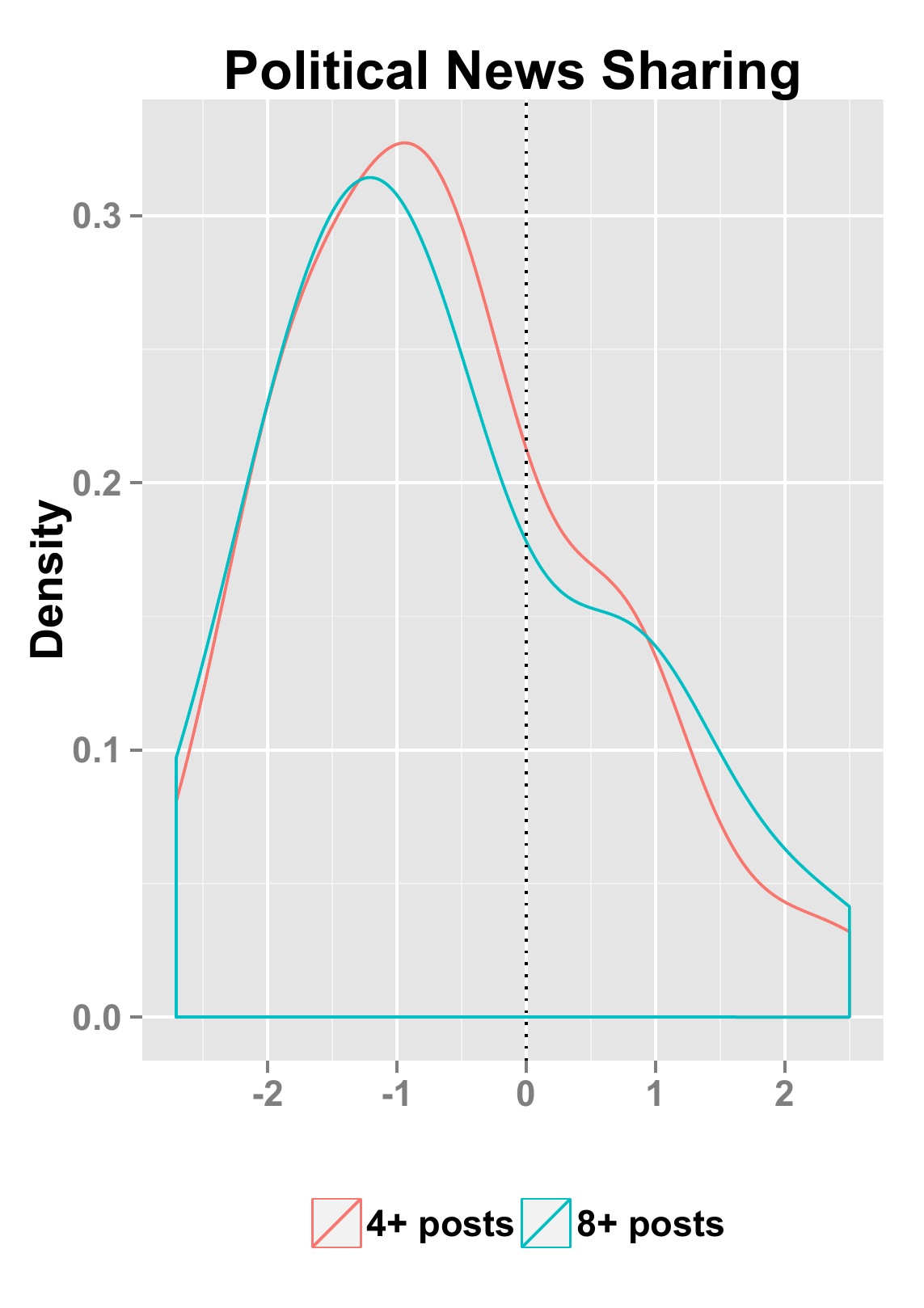} \label{fig:twitter_density_politics_narticleAll} }
\subfigure[Political news (Partisan)]{\includegraphics[width=.23\textwidth]{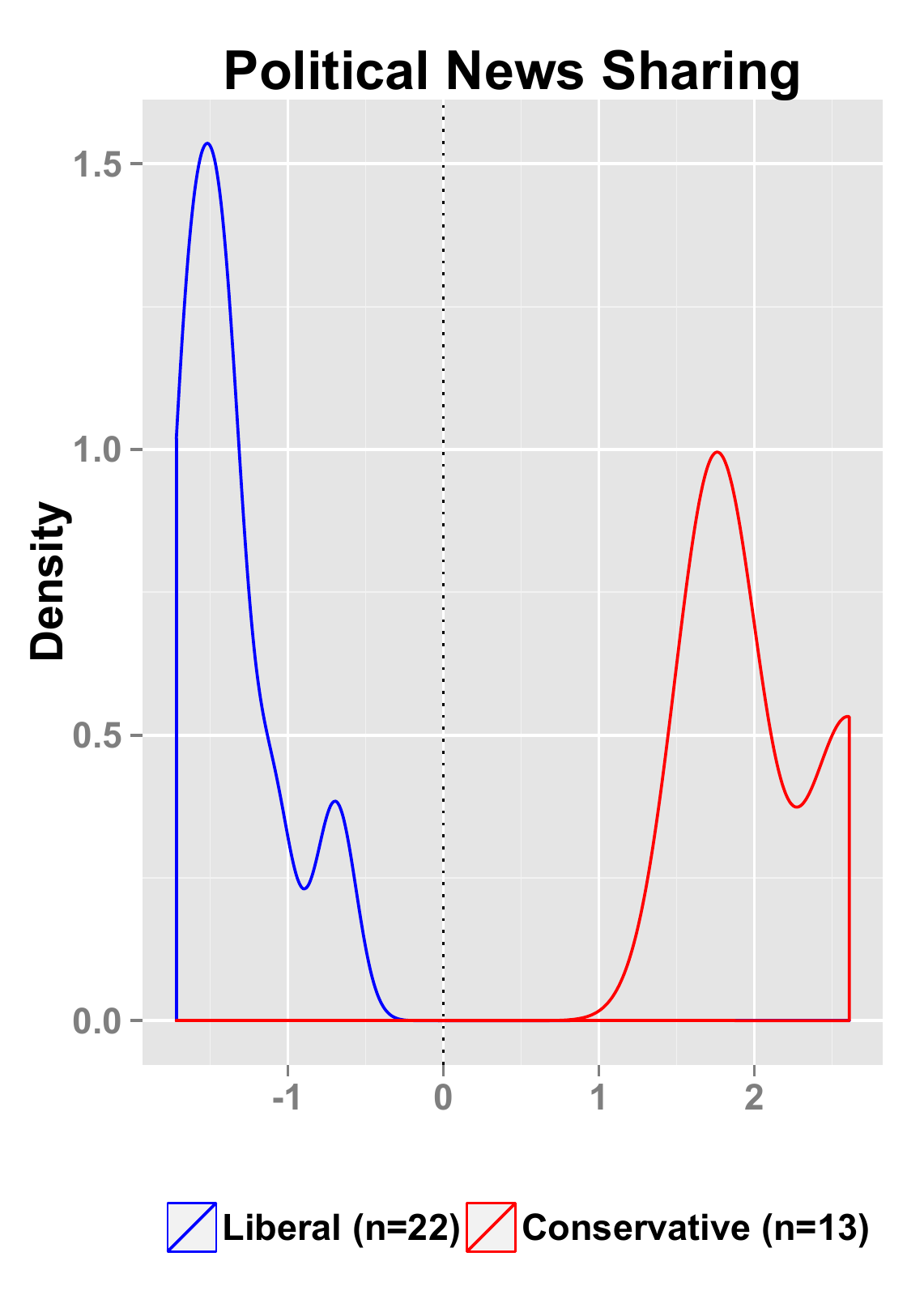} \label{fig:twitter_density_politics_narticleAll_partisan}}
\subfigure[Non-political news (All)]{\includegraphics[width=.23\textwidth]{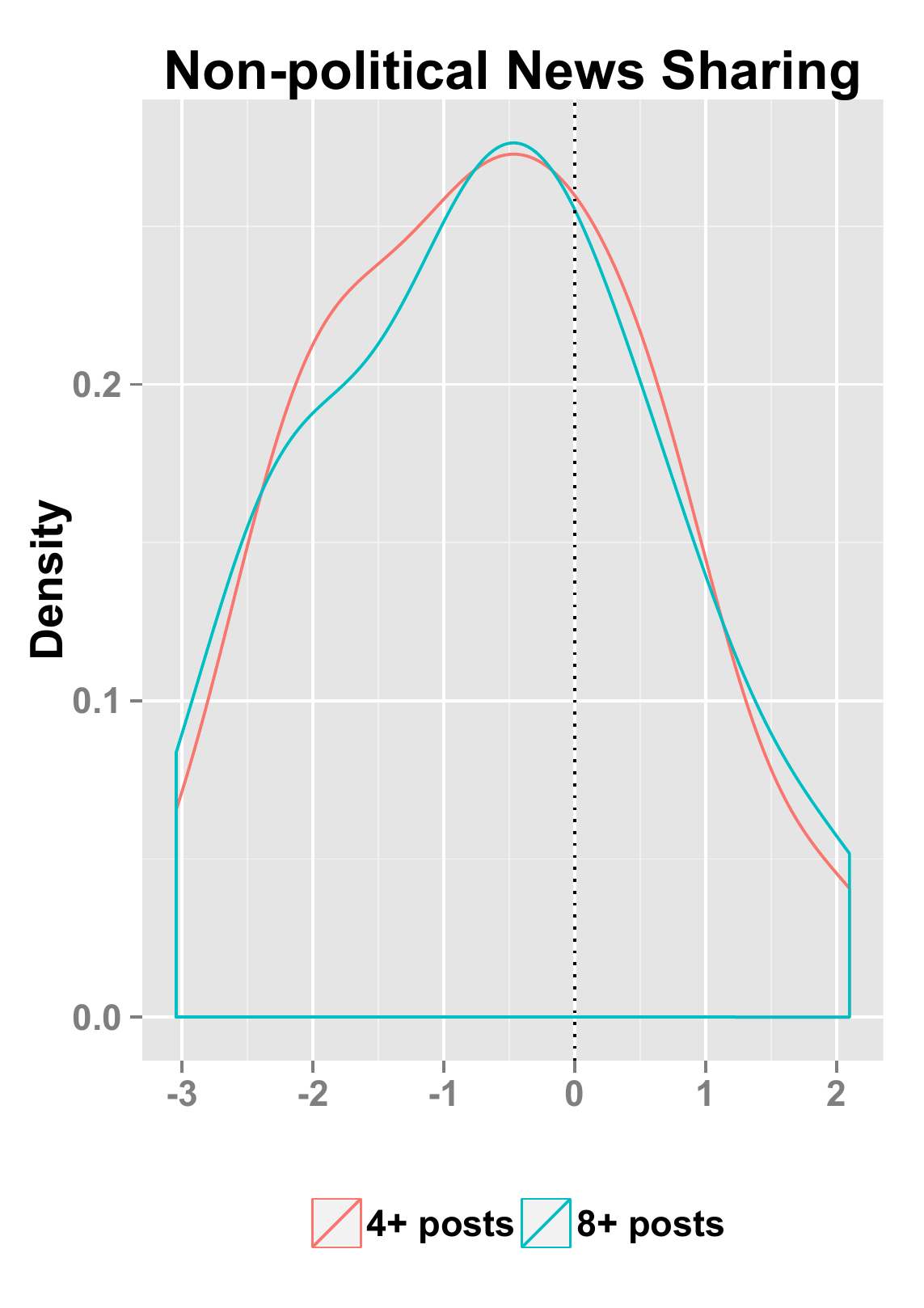} \label{fig:twitter_density_other_narticleAll} }
\subfigure[Non-political news (Partisan)]{\includegraphics[width=.23\textwidth]{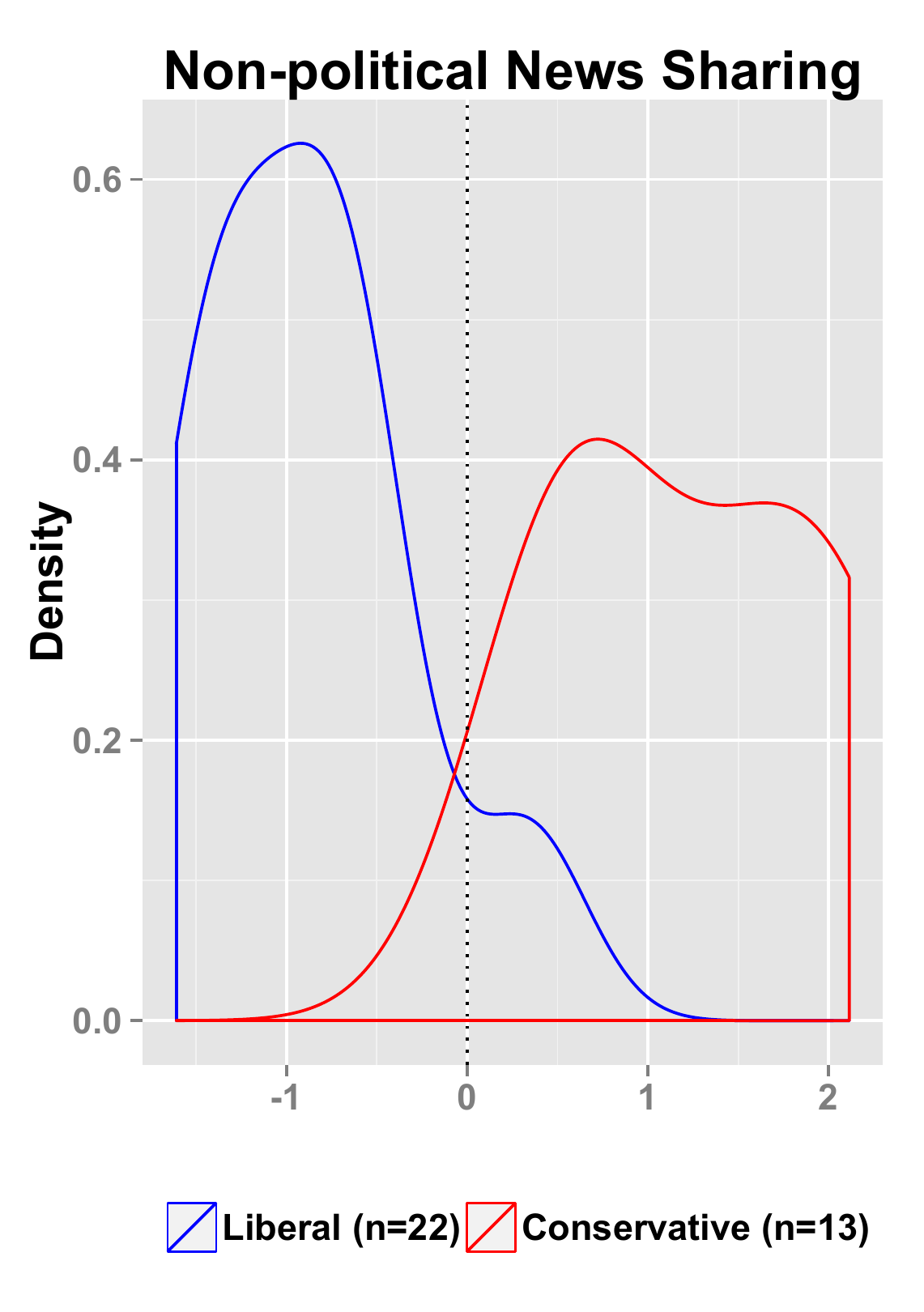} \label{fig:twitter_density_other_narticleAll_partisan}}
\end{center} \vspace*{-6mm}
\caption{\textbf{Net Partisan Skew of political news in Twitter}: \emph{(a}) The majority of our Twitter users are liberal; \emph{(b}) Liberals and conservatives share like-minded news; \emph{(c}) Both share soft news from outlets of no particular leaning (peak is centered around -0.5); \emph{(d}) Partisanship for soft news.} 
\vspace*{-4mm}\label{fig:kdf_net_skew_politics_twitter} \end{figure*}

\subsection{Validation}

\noindent The goal of building this platform and integrating it with Twitter is to test whether partisan sharing has any read-world impact. Before meeting this goal, we need to verify whether the hypotheses that hold on Facebook also hold on Twitter. From the Twitter data, we are able to verify the first two sets of hypotheses (i.e., \emph{H1} and \emph{H2}).

\mbox{} \\
\textbf{Existence.} 

\noindent \emph{[H1] Individuals' news sharing in social media is not balanced but suffers from partisan sharing.} 

To test \emph{[H1]}, we plot the distribution of net partisan skew of Twitter users. There is no bimodal distribution as in the case of Facebook (Figure~\ref{fig:twitter_density_politics_narticleAll}). There are two possible explanations: 1) Twitter itself is known to predominantly liberal~\cite{pewresearch-info-2012}; or 2) conservative users in Twitter share also liberal articles. However, if we consider only partisan users, the bimodal distribution (reflecting the existence of partisan sharing) is back in the picture (Figure~\ref{fig:twitter_density_politics_narticleAll_partisan}).

\mbox{} \\
\textbf{Changes across individuals.} 

\noindent \emph{An individual's level of partisan sharing depends on: \emph{[H2.1] political leaning} and \emph{[H2.2] amount of news shared}.} 

\begin{figure}[t!] \begin{center}
\subfigure[Party]{ \includegraphics[width=.19\textwidth]{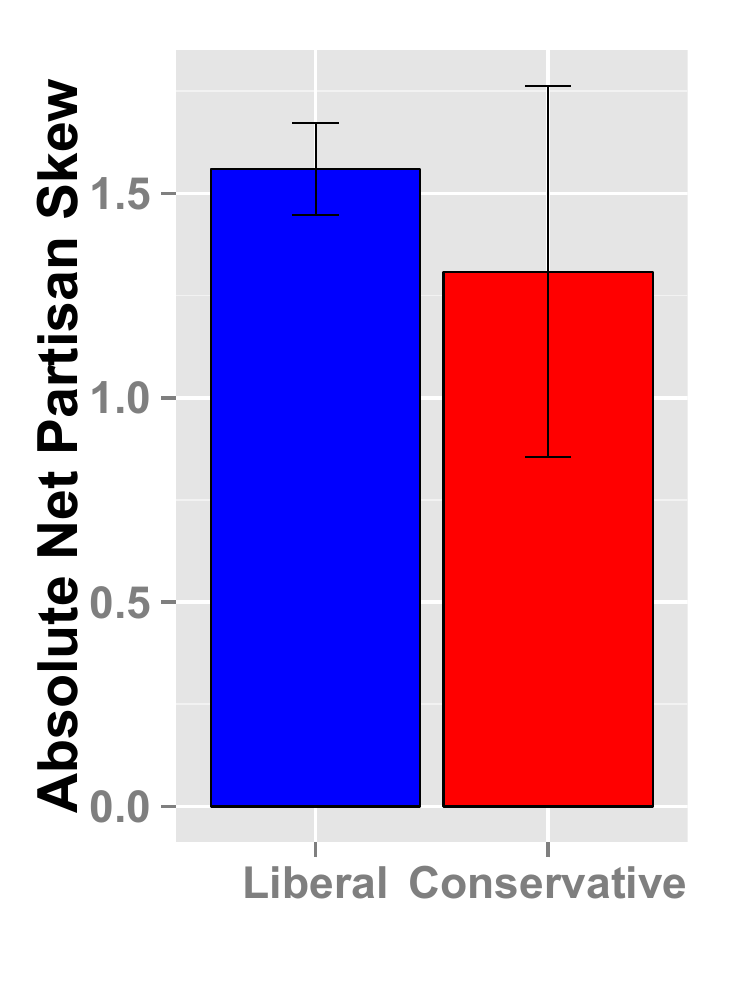} \label{fig:bar_netskew_partisan_twitter}}
\subfigure[Activity]{\includegraphics[width=.26\textwidth]{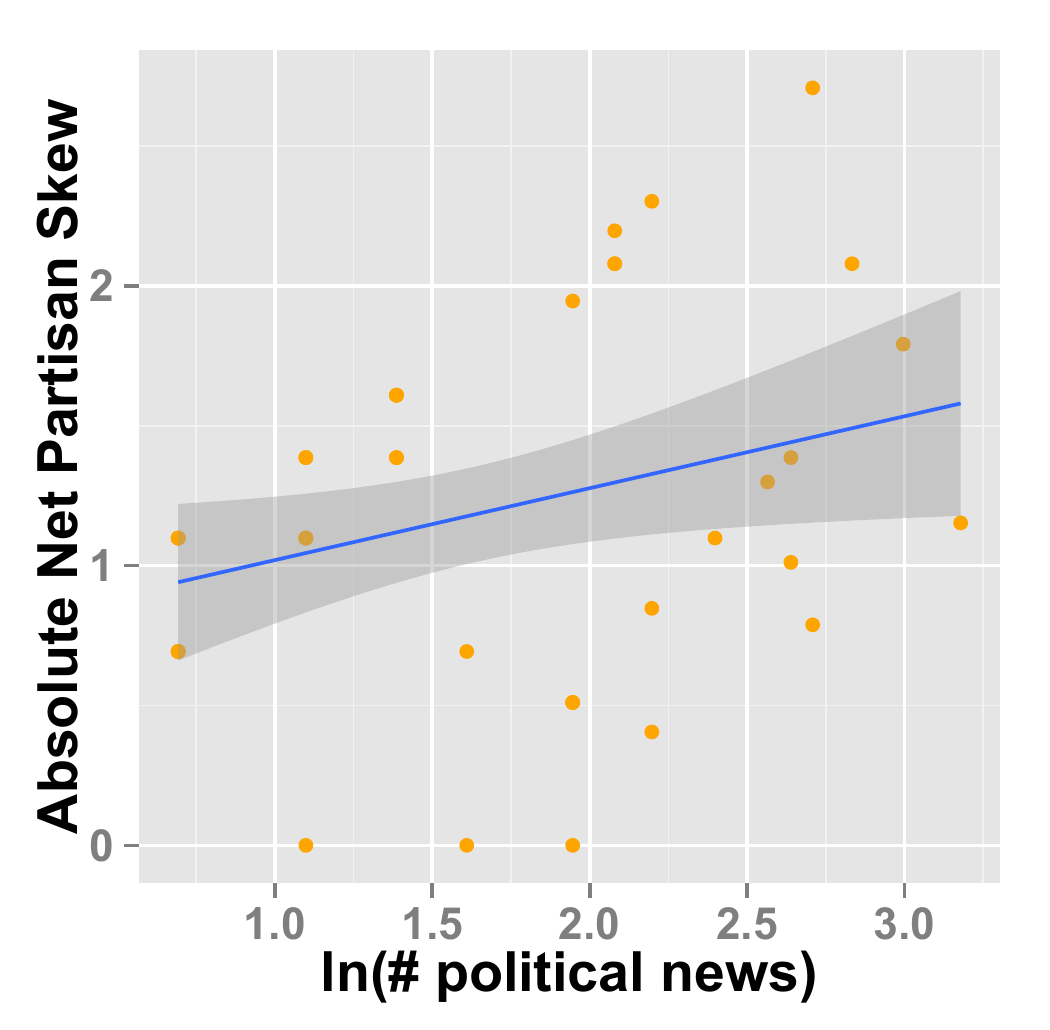} \label{fig:dot_netskew_numpoliticalnews_twitter}}
\end{center} \vspace*{-6mm}
\caption{Net Partisan Skew by \emph{(a)} Party and \emph{(b)} Activity. The absolute value of partisan skew is for UK Twitter users who have shared 4+ articles.}
\vspace*{-2mm}\label{fig:netskew_twitter} \end{figure}

First, to test \emph{[H2.1]}, we compare absolute net partisan skew of liberal users to that of conservative users (Figure~\ref{fig:bar_netskew_partisan_twitter}). The result confirms our previous observation: liberals tend to be more partisan (with net skew of 1.5).  When we examine how absolute net partisan skew varies by amount of political news shared (\emph{[H2.2]}), we find a positive correlation between these two variables (Figure~\ref{fig:dot_netskew_numpoliticalnews_twitter}) with a Pearson's correlation coefficients of $r = .31$ ($p<0.005$) and with a Spearman's correlation coefficients of of $r = .30$ ($p<0.01$), indicating that those who share more are the ones with stronger partisanship.

\subsection{Additional hypotheses}
\label{sect:se_consequence}

\noindent Having ascertained the existence of partisan sharing also on Twitter, we now determine whether it is associated with one's polarized political attitude by administering a survey to the participants of our TV experiment.


\noindent \textit{[H5]} Partisan sharing will relate to polarized political attitudes and, as such, affects one's:
\begin{itemize}[leftmargin=*,noitemsep]
\item[] \textit{[H5.1]} perceived political bias of news outlets;
\item[] \textit{[H5.2]} political knowledge;
\item[] \textit{[H5.3]} voting probability.
\end{itemize}

\begin{figure}[t!] \begin{center}
\subfigure[Liberal]{\includegraphics[width=.22\textwidth]{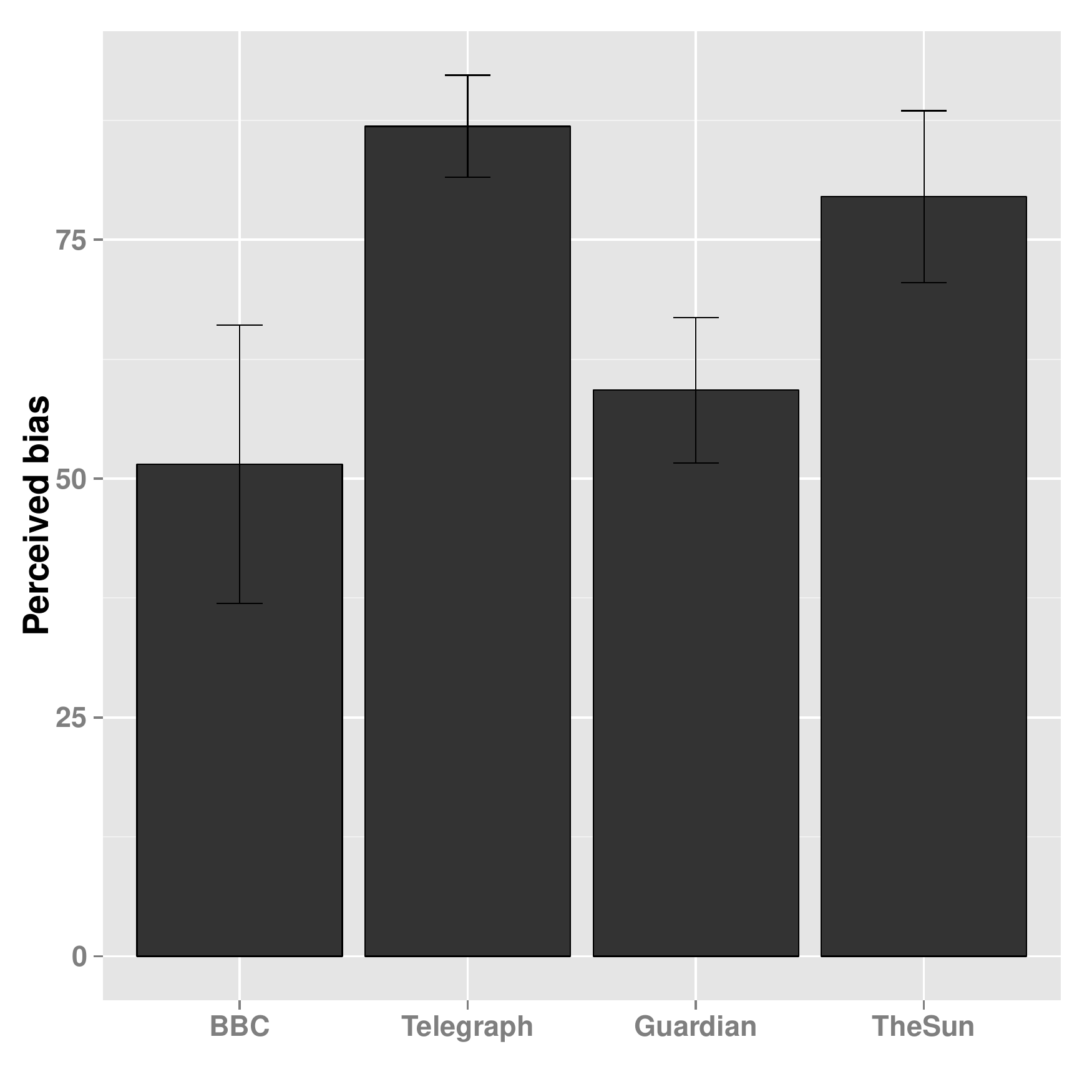}}
\subfigure[Conservative]{\includegraphics[width=.22\textwidth]{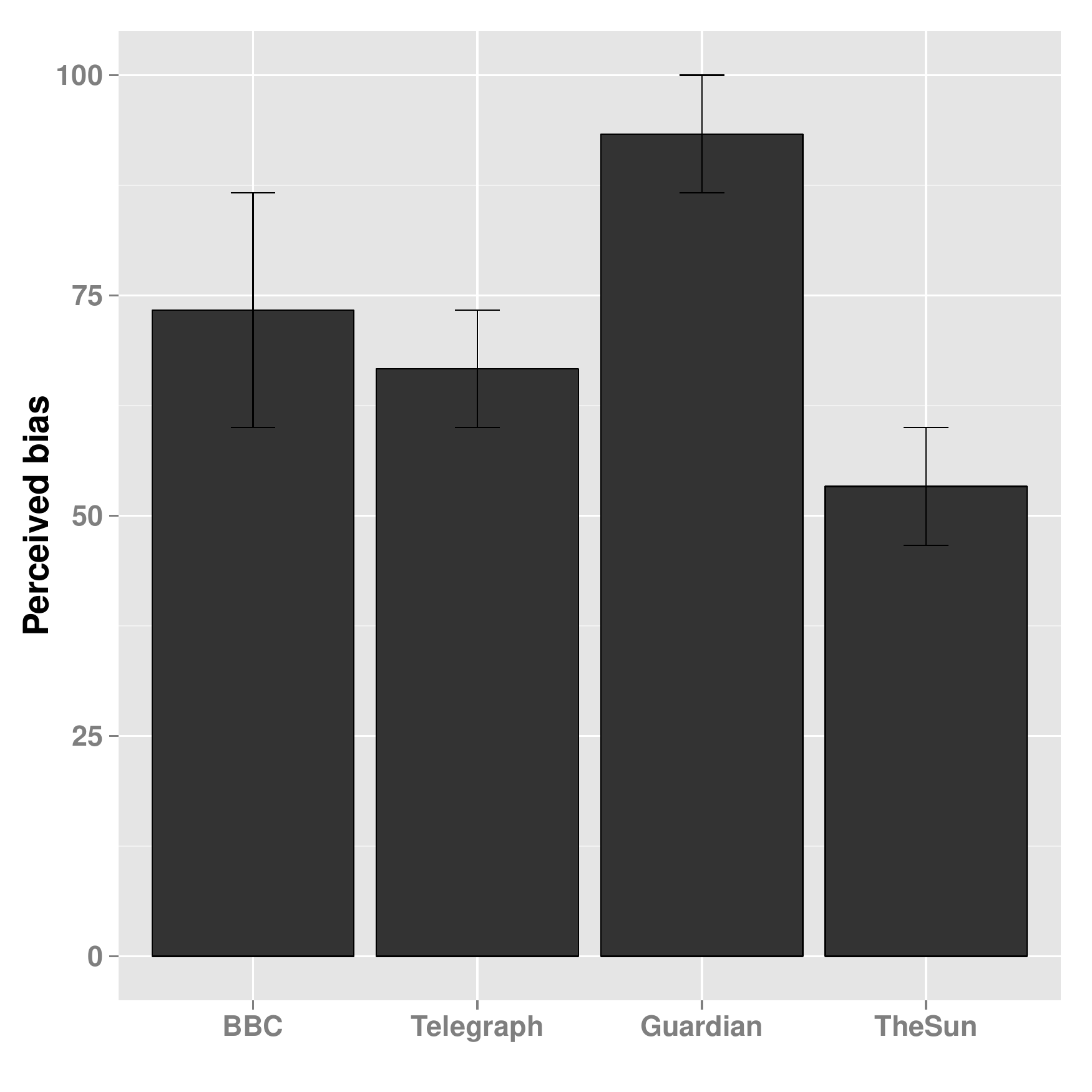}}
\end{center} \vspace*{-6mm}
\caption{Perceived Bias of News Outlets by Party.}
\vspace*{-1mm}\label{fig:perceived_biased} \end{figure}

To test the relationship between perceived bias and partisanship (\emph{H5.1}), we ask our application survey's respondents to which extent they thought four news outlets -- BBC, Telegraph, Guardian, and The Sun -- were  politically biased (the score ranges from 0 to 100,  where 0 means `neutral' and 100 means `strongly biased'). We then compare perceived bias by users with different political leanings. We find that liberal and conservative users significantly differ in their perceptions of the media's leanings (Figure~\ref{fig:perceived_biased}). For example, liberals  perceive the Guardian to be far less biased (56 on a [0,100] scale) than conservatives do (93).  People need to objectively recognize biased reporting to  discount it. The problem is that they are not able to do so: they scrutinize hostile news outlets (those holding views different to their own), while they turn a blind (cognitive) eye to ``friendly'' news outlets.   The ominous consequence of all this is that like-minded information is often perceived to be unbiased and is thus accepted with little scrutiny.

To test the relationship between  political knowledge and partisan sharing (\emph{H5.2}), we need to test one's knowledge. The users of our application administer to a survey. The survey contains 11 questions, 4 of which form together a small  political knowledge quiz about general UK political facts: \emph{Which position is now held by George Osborne?  Is the Queen above the law?  When does the House of Commons scrutinise the government?  Which party has the most members in the House of Commons?} 35 Twitter users answered the survey and the quiz. Given the low number of people, the results need to be taken with caution, but we will see that they confirm previous evidence about news consumption offline, and that deviations from average (error bars) are limited.

\begin{figure}[t!] \begin{center}
\subfigure[Political knowledge]{ \includegraphics[width=.17\textwidth]{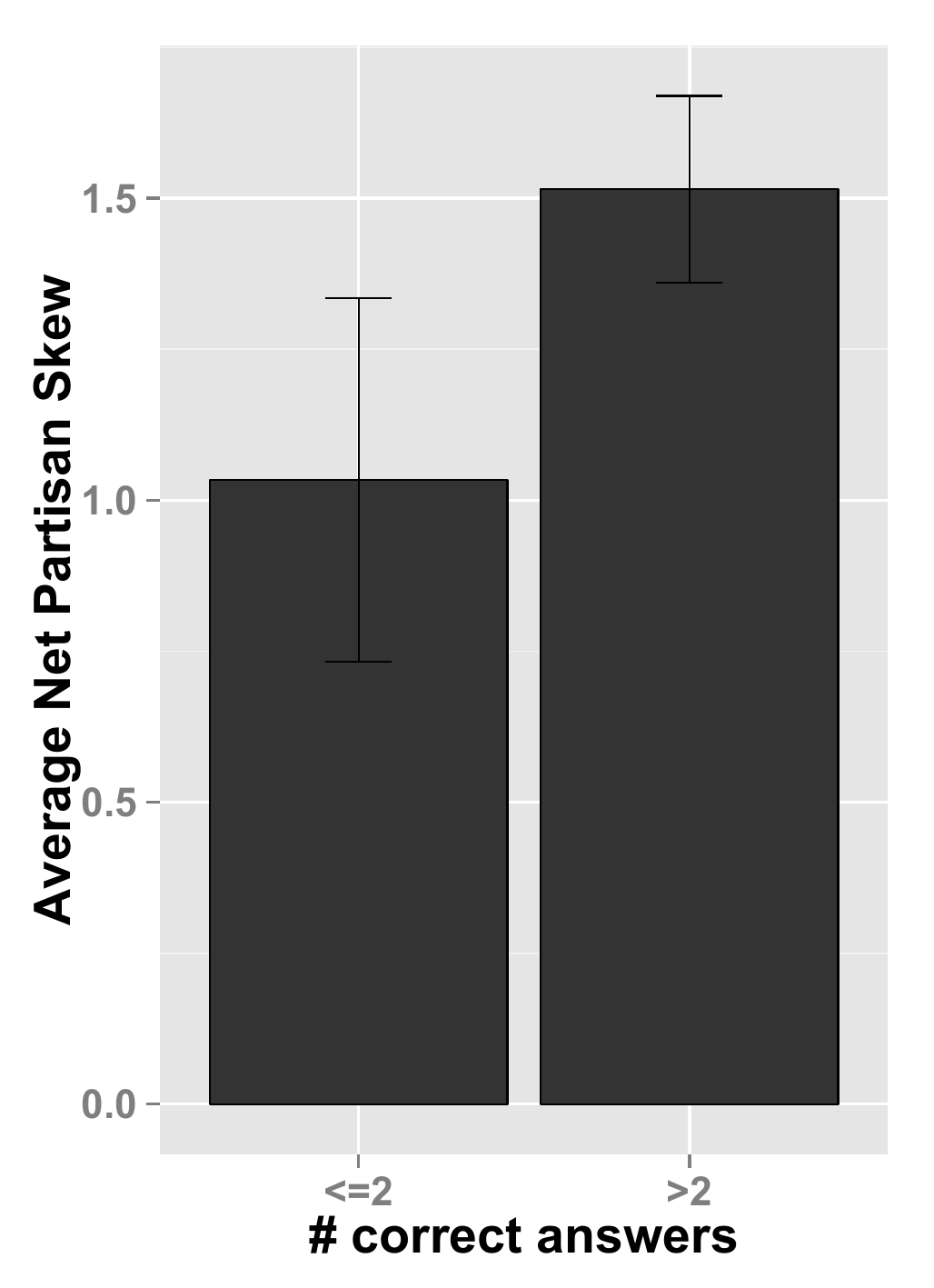} \label{fig:bar_netpartisanskew_pknowledge} }
\subfigure[Voting probability]{\includegraphics[width=.17\textwidth]{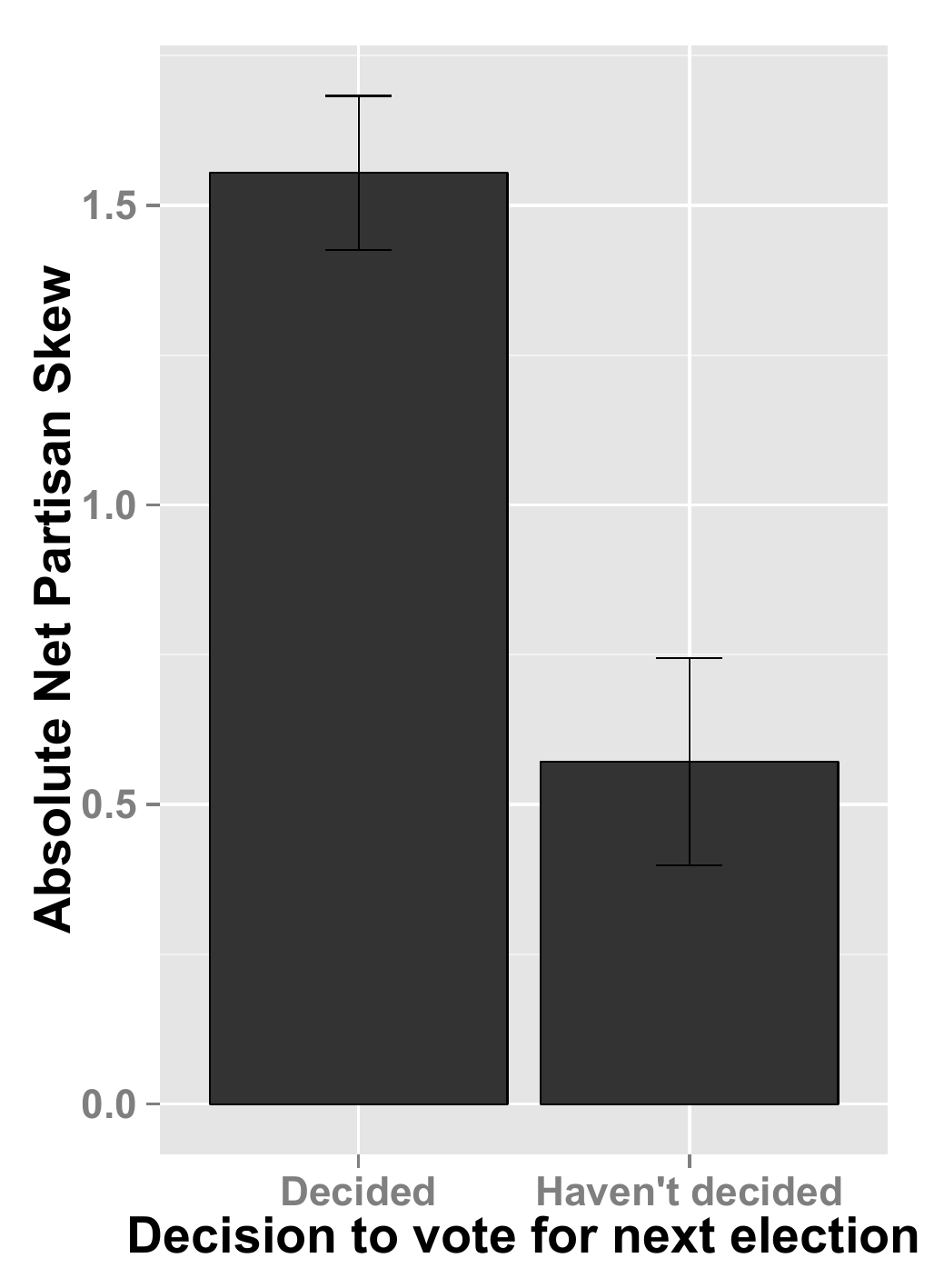} \label{fig:vote_netpartisanskew}}
\end{center} \vspace*{-6mm}
\caption{
Net Partisan Skew by \emph{(a)} Political Knowledge and \emph{(b)} Probability of Voting. The absolute value of partisan skew is for UK Twitter users who have shared 4+ articles.}
\vspace*{-1mm}\label{fig:netskew_twitter_new} \end{figure}

We anticipate that political knowledge (number of correct answers) would be related to partisan sharing. That is because Stroud anticipated that ``partisan selective exposure enhances political knowledge''~\cite{stroud2011}. The results of net partisan skew against number of correct answers in the quiz are shown in  Figure~\ref{fig:bar_netpartisanskew_pknowledge}. As expected, the politically knowledgeable tended to be more partisan than those less knowledgeable.

Finally, to test the relationship between voting probability and partisan sharing (\emph{H5.3}), we contrast two types of Twitter users -- those who have declared to have reached a decision about whether they will vote at the next UK general election in our survey, and those who remain undecided. In line with previous findings in USA~\cite{stroud2011}, UK people who have decided whether to vote are also more partisan than those who remain undecided ($t$(4.558) = 4.566,  $p<0.01$). Despite the small Twitter sample size, the difference is enough not to leave any room for alternative interpretations.

Having their decisions about whether they will vote at the next UK general election, we run a linear regression that predicts one's partisanship based on voting probability:
$$ |net partisan skew| = \alpha + \beta_1 \rm{\textsl{voting}}_\textsl{GeneralElection} $$ The regression has an adjusted $R^2$ of 0.49 and the beta coefficient of $\rm{\textsl{voting}}_\textsl{GeneralElection}$ is 0.98 ($p<0.005$). This result indicates that individual's partisanship could be predicted only by whether they have decided to vote or not.

\subsection{Summary}
\noindent After having ascertained the existence of partisan sharing, we studied the real-world impact of it. As one expects, it is negative as it is strongly related to distorted perceptions of which news outlets are politically biased and which not. However, it is also positive: it is associated with people who are knowledgeable about politics and are actively engaged in political life.

\section{Predicting partisanship}

\noindent Not everyone values diversity~\cite{resnick@chi2010}. 
Thus, to build tools that counter partisan sharing, one would need to identify partisan users first, so that they can adopt a personalized strategy. Several studies have attempted to predict the political leanings of users in SNSs, particularly by their social networks~\cite{PoliticalLeaningOnTwitter} and by the usage of political hashtag in Twitter~\cite{conover@socialcom2010}. Unlike previous work, we attempt to predict a level of partisanship with our Facebook and Twitter datasets.

\subsection{A case study of Facebook}

\noindent Since we have rich demographic data for our Facebook users, we will try to predict their levels of partisanship. More specifically, we consider. The following predictors: 
\begin{itemize}
\item Three Facebook variables: number of Facebook friends, number of postings, and number of likes received from their social contacts.
\item Three personal attributes: sex, age, and size of the city  s(he) lives in.
\item Five personality traits: for our users, we have data from the five-factor model of personality, or the big five, which is the most comprehensive, reliable and useful set of personality concepts~\cite{costa2005,goldberg2006}. An individual is associated with five scores that correspond to the five main personality traits and that form the acronym of OCEAN. Imaginative, spontaneous, and adventurous individuals are high in $Openness$. Individuals who are ambitious, resourceful and persistent individuals are high in $Conscientiousness$. Individuals who are sociable and tend to seek excitement are high in $Extraversion$. Those high in $Agreeableness$ are trusting, altruistic, tender-minded, and are motivated to maintain positive relationships with others. Finally, emotionally liable and impulsive individuals are high in $Neuroticism$. 
\end{itemize}

All predictors undergo a logarithmic transformation, when necessary (e.g., when they are skewed) and are then correlate with net partisan skew. We find that conservatives (high in net partisan skew) tend to be older ($r=0.24$), have less likes from friends ($r=-0.13$),  live in smaller town than liberals ($r=-0.15$), are more emotionally stable and less spontaneous than liberals ($\rm{\textsl{r}}_\textsl{Openness}=-0.20$, $\rm{\textsl{r}}_\textsl{Conscientiousness}=-0.15$, and $\rm{\textsl{r}}_\textsl{Neuroticism}=-0.21$)  All coefficients are statistically significant at level $p$ $<$ 0.005. Next, we study how these predictors are correlated with partisanship. To this end, we correlate them with the absolute value of net partisan skew - the higher his/her absolute value, the more partisan a user.  Out of the eleven predictors, none of them was correlated for conservatives, while only sex was correlated for liberals ($r=0.30$), suggesting that liberal men tend to be more partisan than liberal women. To sum up, it turns out that predicting political leaning (the area which most existing research in computer science has gone into) is far easier than predicting partisanship, which appears to be quite challenging. As a result, it might be very difficult to create tools that effectively counter partisan sharing without being able to identify partisan users.

\subsection{A case study of Twitter}

\noindent The relationship between partisanship and perceived bias also suggests an interesting practical application: knowing how an individual perceives four news outlets to be biased, one could potentially predict the individual's political leaning. To test this, we ask our application survey's respondents to report their  partisanship on a scale [0,100],  where 0 is Labour, 25 is Liberal, 75 is conservative, and 100 is British National Party (BNP) or UK Independence Party (UKIP).\footnote{BNP is far-right political party and UKIP is known to right-wing populist political party in UK.}

Having their views on how they perceived  the four outlets to be biased, we run a linear regression that predicts one's partisanship based on perceived biases\footnote{Given that Twitter users were recruited through web application, connectivity among them was very low (a probability a user is following another was 4\%), allowing us to apply a linear regression model.}:
$$ partisanship = \alpha + \beta_1 \rm{\textsl{bias}}_\textsl{BBC} + \beta_2 \rm{\textsl{bias}}_\textsl{Telegraph} + $$ $$\beta_3 \rm{\textsl{bias}}_\textsl{Guardian}  + \beta_4 \rm{\textsl{bias}}_\textsl{TheSun} $$ The regression has an adjusted $R^2$ of 0.44, which means that as much as 44\% of the variability of an individual's partisanship is explained only by how the individual perceives four news outlets to be biased. The strongest beta coefficients are registered for the left-leaning The Guardian ( $\rm{\textsl{bias}}_\textsl{Guardian}$ =  0.62) and right-leaning The Sun ($\rm{\textsl{bias}}_\textsl{TheSun}$=-0.61).  The different signs suggest that  what differentiates conservatives from  liberals is how they perceive The Guardian and The Sun: the two groups will perceive biased (reliable) a different news outlet. The best way to quickly assess whether one is conservative or not would be to ask him/her how biased The Guardian is and how reliable The Sun is. Indeed, knowing the perceived biases for these two outlets, one could predict partisanship with an adjusted $R^2$ of 0.41. The remaining coefficients are of less importance: $\rm{\textsl{bias}}_\textsl{Telegraph}$=-0.17 (right-leaning newspaper considered biased by left-leaning people); and $\rm{\textsl{bias}}_\textsl{BBC}$=  0.11   (neutral news media corporation considered biased by right-leaning people). All coefficients are statistically significant at level $p<0.01$.

\newpage
\section{Discussion}
\label{sect:discussion}

\vspace*{1mm}
\noindent \textbf{Validating media slant.} To categorize news outlets, we have used four measures of media slant. Had only one been used, we might have been unsure whether our results hold true in general, or whether they are the product of classification artifacts. The four measures are: \textit{scheme1:} \url{http://left-right.org} ; \textit{scheme2:} crowdsourcing platform  \url{http://mondotime.com};  \textit{scheme3:} classification based on scandals~\cite{Larcinese2010} ; and \textit{scheme4:} classification based on congressmen's speeches~\cite{shapiro2010}. All the results we have presented hold for all the four scheme. That is largely because, for any pair of schemes, the two tend to be in agreement. To show this, we consider all possible unordered scheme pairs, and compute their agreement.  By agreement, we mean the number of concordant classifications of the two schemes divided by the total number of classifications. Table~\ref{tab:accuracy}  shows that agreement scores are above 80\%, suggesting that the four schemes are all likely to return very similar classification of media slant.

\vspace*{-2mm}
\begin{table}[h!]
\begin{center}
\small \frenchspacing
\hspace*{-5mm}
\begin{tabular}{c|rrrr}
\textbf{Agreement} & \textit{scheme1} & \textit{scheme2} & \textit{scheme3} & \textit{scheme1} \\
\hline
\textit{scheme1} & 100 & \textcolor{gray}{93.67}  & \textcolor{gray}{81.48} & \textcolor{gray}{100}\\
\textit{scheme2} & 93.67 & 100 & \textcolor{gray}{81.48} & \textcolor{gray}{95.83}\\
\textit{scheme3} & 81.48 & 81.48 & 100 & \textcolor{gray}{100}\\
\textit{scheme4} & 100 & 95.83 & 100 & 100\\
\end{tabular}
\end{center}
\vspace*{-3mm}
\caption{Pairwise-Agreement of Media Slant Measures.}
\vspace*{-4mm}
\label{tab:accuracy}
\end{table}

\begin{figure}[t!] \begin{center}
\includegraphics[width=.42\textwidth]{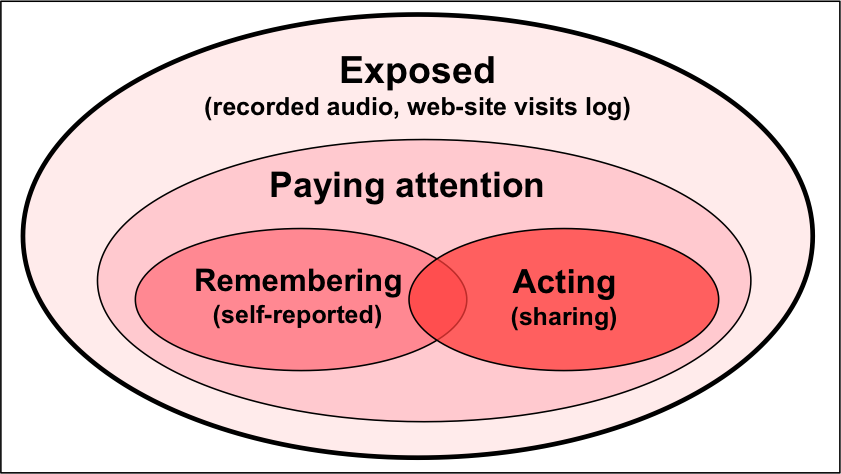}
\end{center} \vspace*{-6mm}
\caption{Attention Dynamics.}
\vspace*{-3mm}\label{fig:level_of_perception} \end{figure}

\vspace*{1.5mm}
\noindent \textbf{Being exposed does not necessarily translate into actions.}  Previous work that supports selective exposure has mostly measured news consumption from self-reported survey data. The little work that has gone into the direct and unobtrusive measurement of news consumption has conflated active with passive exposure. By active exposure, we refer to a situation in which one is either paying attention to news to which (s)he is exposed and, eventually, is translating that attention into action (e.g., calling a friend to chat about the latest political scandal) or is able to recall (e.g., telling an interviewer about the latest political scandal one has read). Instead, by passive exposure, we refer to a situation in which one is exposed to news (s)he is not paying attention to (because, e.g.,  (s)he is multi-tasking). In this work, we have measured how people act upon a piece of news by analyzing what Facebook users actually \emph{share} with their social contacts. Our results are in line with those produced upon self-reported data~\cite{stroud2011} and are in contrast with studies of direct measurements~\cite{shapiro2011, lacour2012}. One possible explanation is that direct measurements capture exposure to news but not necessarily attention (Figure~\ref{fig:level_of_perception}) -- one might be exposed to a piece of news without paying attention to it. By contrast, telling to have read a piece of news  or posting it on Facebook  can only happen if one has paid attention to the piece of news in the first place. By studying news sharing, we have moved the literature forward by measuring active (as opposed to passive) exposure and found evidence for partisan sharing.

\mbox{ } \\
\noindent \textbf{Limitations.} This work has some main limitations. First, our Twitter and Facebook users represent a specific subgroup. Our Twitter users are individuals who are definitely interested in politics as they watch Question Time and tweet about it. Since cultural guidelines clearly exist about political behavior and attitude, one should best consider that our results are likely to hold for that specific group of Twitter users. Similarly, we do not attempt to generalize our Facebook results -- they apply to Facebook users who live in the United States. Secondly, the sample size of Twitter users is limited. As such, the results should only be considered to be preliminary. Yet, they seem to be reasonable for two main reasons: 1) error bars allow us to distinguish which results are more definite than others; 2) results on Twitter and Facebook are consistent, and that speaks to their \emph{external validity}: it is no coincidence that two different platforms in two different countries show similar results. The third limitation of our work is that we do not have any data on why a user shares an article. If a user share a news article of an hostile media outlet, it  does not necessarily mean that (s)he is vouching for it -- (s)he might simply make fun of it. However, given the large sample size on Facebook, such an effect would be likely randomized. The fourth limitation of our work is that the assumption that articles published by a news outlet matches the outlet's political slant. This is reasonable based on the literature of political science, which suggests that even a factual article, as opposed to an op-ed (opinion-editorial) article, often follows a political slant of its source. The last limitation is that we could not test causality for our hypotheses as we do not hold enough data.

\vspace*{1.5mm}
\noindent \textbf{Engaging undecided voters through social media.} Social media have been used by US political campaigners to engage the public. Since past elections have been determined by independent and undecided voters, ``especially those women voters who decide late''~\cite{economist2012}, it might be beneficial to identify the undecided, indifferent, procrastinating, and nonparticipating voters. To do so, this work has suggested that one could search for social media users who have shared only a limited number of political news articles before the election. That strategy would directly target undecided voters. More sophisticated  targeting strategies could tap into social influence~\cite{engagement@pewresearch2012}.  One could, for example, identify  users who are both partisan (with a simple computation of their net partisan skew) and have a considerable number of social contacts who are undecided: they are in the best (social networking) position to influence a large fraction of undecided voters.

\vspace*{-3mm}
\section{Conclusion}
\label{sec:conclusion}
\noindent In large part, political views in the United States are formed nowadays by either television or the Internet. In the past, the structure of television news was built around a broader electorate, and that has changed with the introduction of cable TV: people crave like-minded news channels and avoid politically hostile ones. Do online habits keep up with television ones? This question is too important to be left to the unknown, not least because of the recent decline of national networks and newspapers and the rise of ever more online social media in news industry. This is the first study to unobtrusively measure partisan sharing in the context of online news consumption. We have shown that partisan sharing still exists and  does depend on a variety of factors. Consequently, in the near future, as the current structure of online media consolidates, we might be left with a political discourse driven by echo chambers. It is important, then, to create alternative media that brings together left, right, and center. 

\vspace*{1.5mm}
\noindent \textbf{Acknowledgment} Jisun An was supported in part by the Google European Doctoral Fellowship in Social Computing. 

%

\balance

\bibliographystyle{abbrv}

\balancecolumns


%
%

\balancecolumns
\end{document}